\documentclass[fleqn,usenatbib]{mnras}

\usepackage[T1]{fontenc}
\usepackage{ae,aecompl}

\usepackage{graphicx}	
\usepackage{amsmath}	
\usepackage{amssymb}	

\newcommand{\Msun}{M_{\odot}}

\newcommand{\nue}{\nu_{\rm e}}
\newcommand{\nub}{\bar{\nu}_{\rm e}}
\newcommand{\nux}{\nu_{\rm x}}
\newcommand{\Eexp}{E_{\rm exp}}
\newcommand{\Ediag}{E_{\rm diag}}
\newcommand{\Mni}{M_{\rm Ni}}

\title[3D CCSN simulation for SN 1987A binary progenitor]{Three-dimensional simulation of a core-collapse supernova for a binary star progenitor of SN 1987A}

\author[K. Nakamura et al.]{
Ko Nakamura,$^{1,2}$\thanks{E-mail: nakamurako@fukuoka-u.ac.jp}
Tomoya Takiwaki$^{3}$
and Kei Kotake$^{1,2}$
\\
$^{1}$Department of applied physics, Fukuoka University, Nanakuma Jonan 8-19-1, Fukuoka 814-0180, Japan \\
$^{2}$Research Institute of Stellar Explosive Phenomena, Fukuoka University, Nanakuma Jonan 8-19-1, Fukuoka 814-0180, Japan \\
$^{3}$National Astronomical Observatory of Japan, Osawa 2-21-1, Mitaka, Tokyo 181-8588, Japan
}

\date{Accepted XXX. Received YYY; in original form ZZZ}

\pubyear{2022}

\begin{document}
\label{firstpage}
\pagerange{\pageref{firstpage}--\pageref{lastpage}}
\maketitle

\begin{abstract}
We present results from a self-consistent, non-rotating core-collapse supernova simulation in three spatial dimensions 
using  
a binary evolution progenitor model of SN~1987A by Urushibata et al. (2018).
This $18.3\Msun$ progenitor model is evolved from a slow-merger of 14 and $9\Msun$ stars, 
and it satisfies most of the observational constraints such as red-to-blue evolution, lifetime, total mass and position in the Hertzsprung-Russell diagram at collapse, and chemical anomalies. 
Our simulation is initiated from a spherically symmetric collapse and mapped to the three-dimensional coordinates at 10 ms after bounce to follow the non-spherical hydrodynamics evolution. 
We obtain the neutrino-driven shock revival for this progenitor at $\sim$ 350\,ms after bounce, leading to the formation of a newly-born neutron star with average gravitational mass $\sim 1.35\,\Msun$ and spin period $\sim 0.1$ s. 
We also discuss the detectability of gravitational wave and neutrino signals for a Galactic event with the same characteristics as SN~1987A.
At our final simulation time ($\sim 660$\,ms postbounce), the diagnostic explosion energy, though still growing, is smaller (0.15 foe) compared to the observed value (1.5 foe).
The $^{56}$Ni mass obtained from the simulation ($0.01\Msun$) is also smaller than the reported mass from SN~1987A ($0.07\Msun$). 
Long-term simulation including several missing physical ingredients in our three-dimensional models such as rotation, magnetic fields, or more elaborate neutrino opacities should be done to bridge the gap between the theoretical predictions and the observed values.
\end{abstract}




\section{Introduction} \label{sec:intro}

SN 1987A, which emerged in the Large Magellanic Cloud located at a distance of $49.59 \pm 0.63$\,kpc \citep{pietrzynski19}, 
is a landmark event in astrophysics.
Multiwavelength studies of SN 1987A have provided unprecedented details of supernova features 
such as the time evolution of the bolometric luminosity \citep{shigeyama88,shigeyama90,suntzeff90} 
and energy spectrum at all wave bands from infrared to $\gamma$-ray 
\citep[for a review of the observational features of SN 1987A, see, e.g.,][]{arnett89,mccray93,mccray16}. 
Moreover, a supernova neutrino burst was detected by several neutrino detectors such as 
Kamiokande-II \citep{hirata87,hirata88}, 
IMB \citep{bionta1987,bratton88}, 
and Baksan \citep{alekseev87,alexeyev88}. 
Although only about two dozen of the $\sim 10^{28}$ supernova neutrinos that passed through the Earth were detected, 
they provided us with the first (and ever only one) direct evidence of the supernova (SN) driven by the collapsing core of a dying star 
\citep{sato87}.
This detection implied that there had been a proto-neutron star at least for 10 seconds and it declared the dawn of the neutrino astrophysics. 

The nature of the central remnant of SN 1987A
-- the compact remnant left behind the explosion is whether a neutron star or a black hole, for example --
has been a mystery since 
observational searches for the remnant in radio and X-rays have been unsuccessful \citep[e.g.,][]{alp18,esposito18} owing to the dust and the ring surrounding the supernova remnant. 
Recently, however, \citet{cigan19} have presented high angular resolution images of dust and molecules in SN 1987A ejecta obtained from the Atacama Large Millimeter/submillimeter Array (ALMA) and concluded that the presence of a compact source in the remnant is strongly indicated. 
The infrared excess could be due to the decay of isotopes like $^{44}$Ti, accretion luminosity from a neutron star or black hole, magnetospheric emission, a wind originating from the spindown of a pulsar, or thermal emission from an embedded, cooling neutron star. 
\citet{page20} carefully investigated all possibilities and concluded that the cooling NS scenario is the most plausible.

The light curve and spectra indicate that SN 1987A has roughly typical explosion energy 
\citep[$\sim 1.5 \times 10^{51}$\,erg,][]{jerkstrand20} 
and $^{56}$Ni mass \citep[$0.07\,\Msun$,][]{bouchet91,mccray93,seitenzahl14}. 
On the other hand, SN 1987A is a unique supernova in many senses. 
First, a blue hot surface of the progenitor star Sk-69$^\circ$202 found in the pre-explosion images \citep{blanco87,walborn87}, 
as well as the light curve without the typical plateau phase of type II-P SNe \citep{catchpole88,hamuy88} 
and the relatively short period of time delay (three hours) between the neutrino burst detection and the shock breakout emission, 
suggest that the progenitor was a blue-supergiant (BSG) \citep{woosley88,arnett89}. 
Second, this BSG star is 
expected to have been a red-supergiant (RSG) $2 \times 10^4$\,yrs before exploding, since 
there are three ring-like nebulae surrounding the supernova remnant \citep{wampler89,wampler90,burrowsCJ95,france10} 
with the high He and CNO abundance rations \citep{lundqvist96,mattila10} 
and the expansion velocity comparable to the RSG wind velocity.

To explain these anomalous features, many pre-explosion evolution scenarios of the progenitor star 
have been proposed soon after the emergence of SN~1987A 
(see \citealt{Podsiadlowski92b} for a review of the following classical progenitor models): 
extreme-mass-loss models \citep{maeder87,wood87}, 
helium-enrichment models \citep{saio88}, 
low-metallicity models \citep{arnett87,hillebrandt87}, 
rapid-rotation models \citep{weiss88,langer91}, 
and restricted-convection models \citep{woosley88,langer89,weiss89}. 
Also, binary interaction has been considered from early on: 
accretion models \citep{barkat89,podsiadlowski89,deloore92},
companion models \citep{fabian87,joss88},
and merger models \citep{hillebrandt89,podsiadlowski90}. 

The majority of massive stars live in binary or multiple systems 
\citep{duchene13, sana13}, 
and binary interactions can alter both the surface and core structure of the core-collapse supernova (CCSN) progenitor stars 
\citep{laplace21,vartanyan21}. 
One of the possible scenarios to the SN~1987A progenitor is a slow merger of binary stars 
where the stars in a close binary evolve into a common envelope phase 
after which the secondary star is gradually dissolved inside the common envelope 
in a much longer time-scale than the dynamical time-scale of the secondary \citep{podsiadlowski90,Podsiadlowski92a,Ivanova02}. 
Recently new progenitor models based on the slow merger scenario have been constructed \citep{menon17,urushibata18}. 
The models of \citet{menon17} were the first to evolve a binary-merger model 
until core-collapse for SN~1987A’s progenitor and successfully reproduced 
the progenitor characteristics 
such as red-to-blue evolution, 
position in the Hertzsprung-Russell diagram at collapse, 
and the enrichment of helium and nitrogen in the progenitor, 
along with the lifetime of the BSG phase. 
\citet{menon17} employed a simplified effective-merger model, and they did not compute mass loss in the merger phase. In contrast,
\citet{urushibata18} have taken into account the spinning-up of the stellar envelope and subsequent mass loss due to angular momentum transportation from the orbit. 
They obtained enough mass ejection to explain the circumstellar nebula, as well as the characteristics of SN~1987A’s progenitor.

In parallel to the study of the progenitors of SN~1987A, a separate effort has been devoted to understand the explosion characteristics of SN~1987A, including the explosion energy, distribution of synthesized elements, emission of multimessenger signals and properties of the central remnant. 
\citet{kifo06} investigated matter mixing with realistic explosion models 
based on two-dimensional (2D) hydrodynamical simulations of neutrino-driven CCSNe aided by convection and/or SASI. The authors found that a globally aspherical explosion dominated by low-order unstable modes 
with explosion energy of $2 \times 10^{51}$\,erg produces high-velocity $^{56}$Ni clumps 
as inferred from the observed [Fe II] line profiles in the remnant of SN 1987A.
\citet{wongwathanarat15} investigated the dependence of matter mixing on single-star progenitor models based on three-dimensional (3D) hydrodynamic simulations and also obtained such a high-velocity $^{56}$Ni. 
\citet{utrobin15} modelled optical light curves based on 3D hydrodynamic models. 
Among the investigated models, only one BSG model reproduces the dome-like shape of the light-curve maximum of SN 1987A. 
\citet{utrobin19} investigated matter mixing in explosions of a large sample of BSG models 
and concluded that single-star progenitor models have difficulties in reproducing observational constraints.

The binary-merger models of \citet{menon17} were investigated by 1D explosion simulations \citep{menon19}.
In comparison with the available single-star BSG progenitor models, the post-merger models were able to better reproduce the shape of the light curve of SN~1987A along with the light curves of other peculiar Type II supernovae like SN~1987A. 
\citet{utrobin21} performed 3D neutrino-driven explosions 
and followed the evolution of the shock wave long after explosion. 
They succeeded in reproducing the $^{56}$Ni mass and mixing in the ejecta material, leading to a good reproduction of  light curve of SN~1987A. 
Employing the binary progenitor model of \citet{urushibata18} and FLASH code \citep{fryer00}, 
\citet{ono20} performed hydrodynamic simulations in the 3D Cartesian coordinates 
focusing on the matter mixing in SN 1987A. 
They succeeded in reproducing high-velocity $^{56}$Ni inferred from the observed [Fe II] line profiles in the remnant of SN 1987A. 
In all of these simulations based on single-star and binary progenitor models, 
the inner region corresponding to a compact object was excluded from the calculation and a parametric energy or neutrino emission was injected to drive an explosion. 

In this paper, we report the results of our self-consistent 3D CCSN simulation employing the binary-merger progenitor model of \citet{urushibata18}.
We solve hydrodynamic evolution and energy-dependent neutrino transport from the centre to the outer boundary 
at 10,000\,km. 
A number of self-consistent 3D simulations with energy-dependent neutrino transport 
\citep[e.g.][]{takiwaki14,melson15a,melson15b,lentz15,janka16,ott18,summa18,mueller18,mueller19,burrows20,iwakami20,stockinger20,takiwaki21,matsumoto22}
have recently been demonstrated based on the neutrino-driven mechanism 
\citep{colgate66,arnett66,bethe85}.
Our simulations have an advantage in taking account of nuclear network calculation 
so that we can provide an estimate of explosive nucleosynthesis yield.
Furthermore, we investigate in detail the properties of 
the SN~1987A's central remnant and multi-messenger signals obtained from the results of the 3D self-consistent simulation.

This paper is organized as follows. 
Section 2 summarizes our numerical methods and the progenitor model employed in this work. 
We present our 3D results in section 3, which consists of shock evolution (\S 3.1), neutrino emission and PNS convection (\S 3.2), 
and multimessenger signals including gravitational wave (\S 3.3).
We conclude with discussions in section 4.

\section{Numerical setups} \label{sec:setup}

The progenitor model we use is based on the slow merger scenario \citep{podsiadlowski90,Podsiadlowski92a}, which explains the evolution of SN 1987A progenitor from an RSG to a BSG as a result of stellar merger and deflation of the primary RSG in a binary system. 
Recently, similar attempts to model the evolution of the SN 1987A progenitor star, Sk -69$^\circ$202, have been presented \citep{menon17,urushibata18}. 
In particular, \citet{urushibata18} have considered the spinning-up of the stellar envelope 
and their best model ($14+9\Msun$ binary with a particular parameter set of merging) successfully reproduced the progenitor characteristics such as the N/C ratio of the stellar surface and the timing of the blueward evolution of $2 \times 10^4$\,yr before the collapse, as well as the mass ejection into the circumstellar medium. 
Therefore, we employ the $14+9\Msun$ progenitor model (hereafter m14 model) from \citet{urushibata18}. 
This progenitor is studied by \citet{ono20} focusing on a matter mixing in the outer envelope using FLASH code, and we employ the radiation/hydrodynamic code 3DnSNe \citep{takiwaki16,takiwaki18} to study the dynamical evolution of the core of this progenitor model as a central engine of SN 1987A.

\begin{figure*}
  \begin{center}
  \begin{tabular}{cc}
    \includegraphics[width=0.45\linewidth]{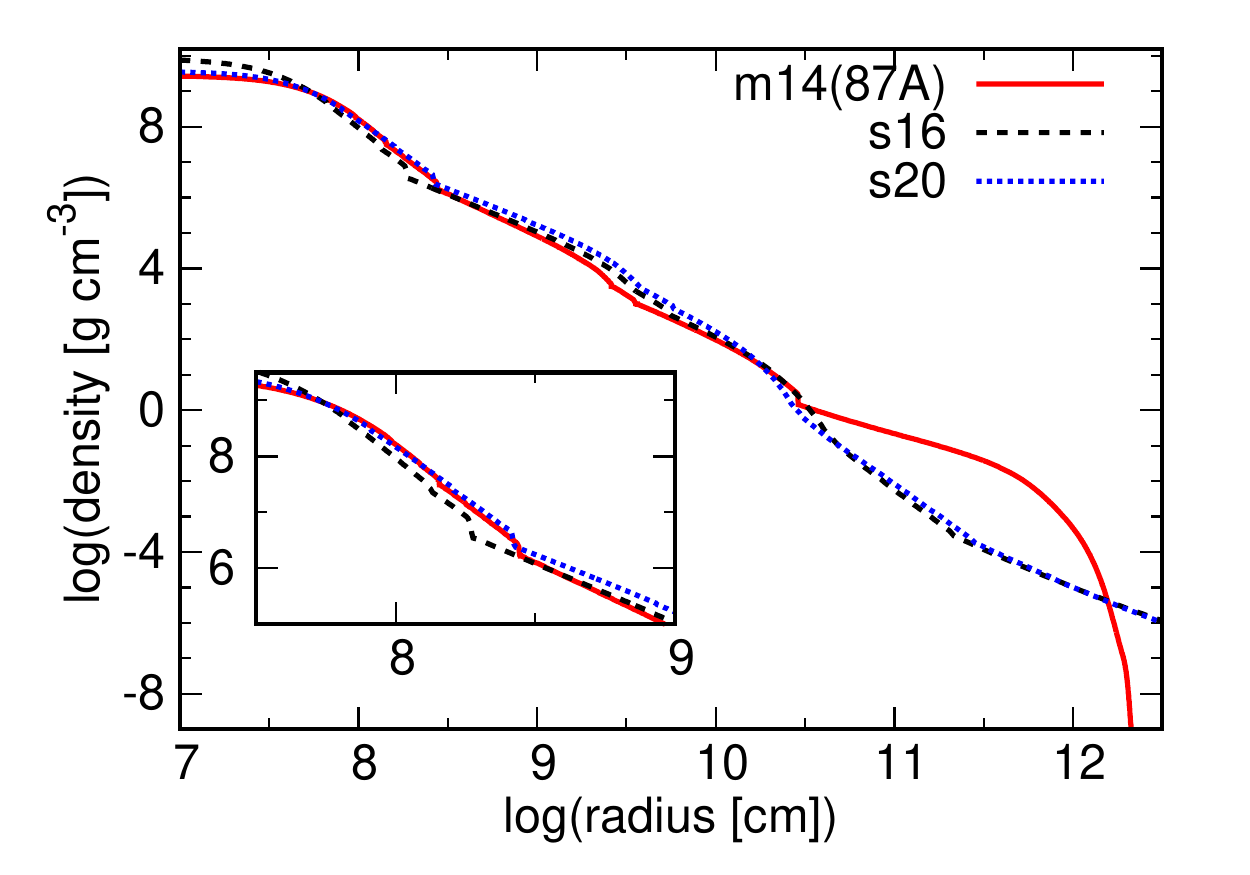} &
    \includegraphics[width=0.45\linewidth]{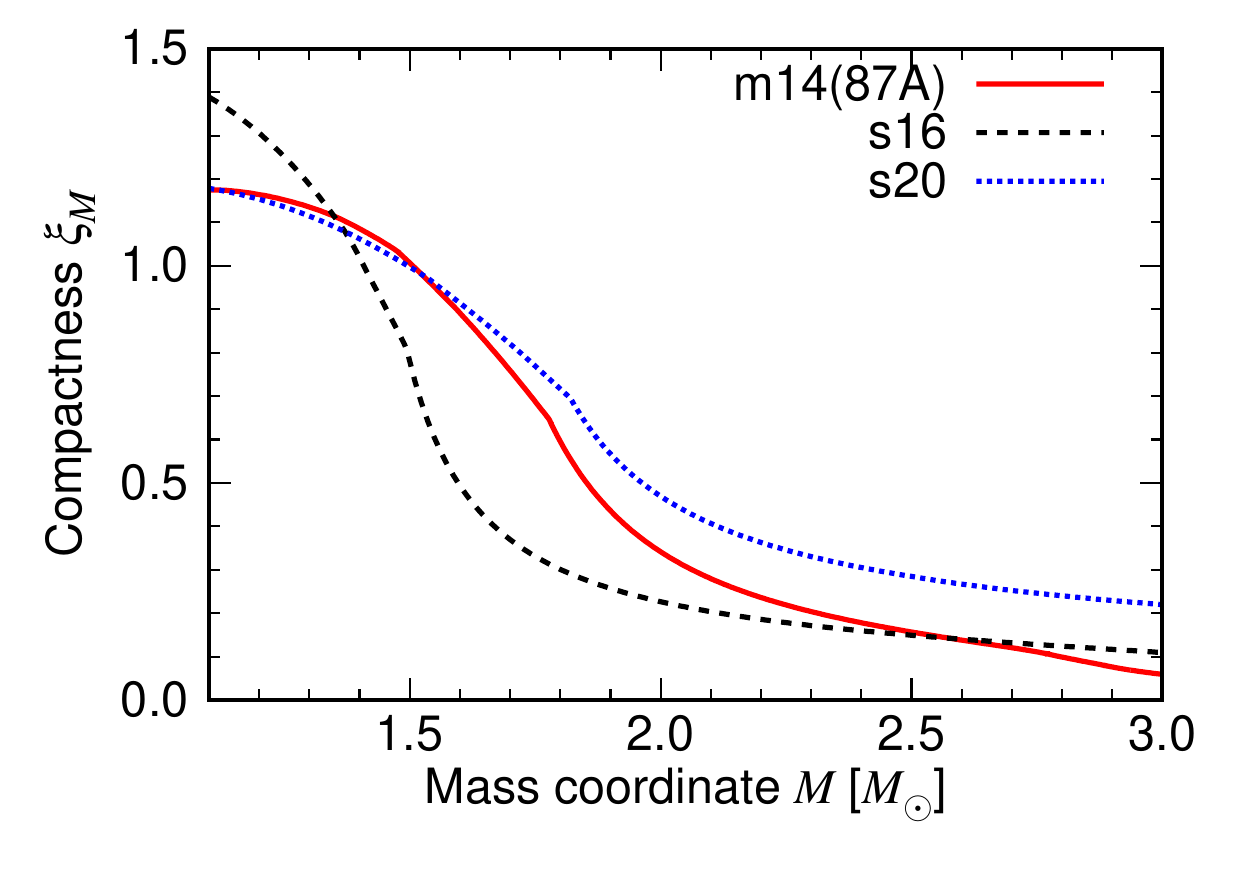} \\
  \end{tabular}
  \end{center}
  \caption{The structure of the SN 1987A progenitor model \citep[m14,][]{urushibata18} compared with 16 and $20\Msun$ pre-SN models from \citet{woosley07}. 
  Left panel: Density profiles of the progenitors. 
  The SN 1987A progenitor is a BSG star and its radius is about one order of magnitude smaller than RSG stars with an extended hydrogen envelope. 
  In the inset, we show the zoomed-in density profile around the iron core surface located at $r \sim 10^{8.3}$ -- $10^{8.5}$ cm. 
  Right panel: the compactness parameter of these three progenitor models as a function of enclosed mass. 
  The compactness of the SN 1987A progenitor is smaller than that of s20 model, which has been reported to explode in 3D simulation \citep{melson15b}.
  }%
  \label{fig:prog}
\end{figure*}

Figure~\ref{fig:prog} describes the structure of the m14 model, compared with 16 and $20\Msun$ pre-SN RSG models from \citet{woosley07}.
There is an outstanding difference in the density profile (left panel) in the outer envelope since the BSG SN 1987A progenitor has a much smaller radius than the other RSG progenitors. 
Note that our simulations are within $10^9$ cm from the center. 
Given that the progenitor's compactness parameter $\xi_M$, which is defined in  \citet{oconnor11} as a function of an enclosed mass $M$, 
\begin{equation}
\xi_M = \frac{M \, [\Msun]}{R(M) \, [1000 \, {\rm km}]},
\end{equation}
is a good diagnostics for the explosion properties. 
The right panel of Figure~\ref{fig:prog} shows the compactness $\xi_M$ as a function of the enclosed mass. 
The m14 model lies between s16 and s20 models, both of which are reported to explode in 3D simulations 
in a self-consistent manner \citep{vartanyan19,melson15b} via neutrino heating, namely without additional input of energy nor increase in the heating rate.

The 3DnSNe code that we employ in this study is a multi-dimensional, three-neutrino-flavour radiation hydrodynamics code constructed to study core-collapse supernovae\footnote{Though we have updated the code to deal with magnetohydrodynamics (MHD) \citep{Matsumoto20}, we shall limit ourselves to report the hydrodynamics results in this work, simply because the 3D MHD runs are still under way.}. 
The neutrino transport is solved by the isotropic diffusion source approximation (IDSA) scheme 
for electron, anti-electron, and heavy lepton neutrinos \citep{idsa,takiwaki16} 
taking the state-of-the-art neutrino opacity \citep{kotake18} 
and discretized neutrino spectrum with 20 energy bins for $0 < \epsilon_\nu \leq 300$\,MeV.
\citet{takiwaki18} have implemented the gravity potential taking account of the effective General Relativistic effect 
\citep[case A in][]{marek06}.

The spatial range of the 1D, 2D, and 3D computational domain is within $r <$10,000\,km in radius and 
divided into 600 non-uniform radial zones. 
Our spatial grid has the finest mesh spacing $dr_{\rm min} = 250$\,m at the centre, and $dr/r$ is better than 1.0\,\% at $r > 100$\,km. 
The 2D model is computed on a spherical coordinate grid with an angular resolution of $n_\theta = 128$ zones. 
For 3D model we simulate with the resolution of $n_\theta \times n_\phi = 64\times 128$ zones. 
Seed perturbations for aspherical instabilities are imposed by hand at 10\,ms after bounce by introducing random perturbations of $0.1\%$ in density on the whole computational grid except for the unshocked core.
Regarding the equation of state (EOS), we use that of \citet{lattimer91} 
with a nuclear incomprehensibility of $K = 220$\,MeV.
At low densities, 
we employ an EOS accounting for photons, electrons, positrons, 
and ideal gas contribution. 
We follow the explosive nucleosynthesis 
by solving a simple nuclear network consisting of 13 alpha-nuclei, 
$^4$He, $^{12}$C, $^{16}$O, $^{20}$Ne, $^{24}$Mg, $^{28}$Si, 
$^{32}$S, $^{36}$Ar, $^{40}$Ca, $^{44}$Ti, $^{48}$Cr, $^{52}$Fe, and $^{56}$Ni.
A feedback from the composition change to the EOS is neglected, 
whereas the energy feedback from the nuclear reactions 
to the hydrodynamic evolution is taken into account as in \citet{nakamura14a}.

Along with these setups, we simulate core collapse, bounce, and subsequent shock evolution 
under the influence of neutrino heating. 
Our simulations are self-consistent 
in the sense that we do not employ any artificial control to drive explosions. 
Note that there is a consensus that a successful supernova explosion 
driven by neutrino heating 
cannot be achieved under the spherical symmetry (1D)
\citep{liebendorfer01,sumiyoshi05}, 
except for a few less massive stellar models (e.g., \citet{kitaura06} for an O-Ne-Mg core and \citet{mori21,nakazato21} for an iron core).
Note also that we simulate the core collapse and bounce in 1D geometry and then map it to the 
3D coordinates at 10 ms after bounce to follow non-spherical evolution, whereas 2D simulation 
is initiated from the beginning of the collapse. 
This difference in the numerical treatment until the core bounce does not play any significant role 
since our simulations start from the spherically symmetric progenitor and non-spherical motions are developed only after the bounce by hydrodynamic instabilities.

\section{Results} \label{sec:3d}
This section reports the results of our 
core-collapse simulations
for the latest SN 1987A progenitor model, 
primarily focusing on the 3D simulation. 
Our simulations are based on the neutrino heating mechanism and failed in explosion for 
the 1D spherically symmetric case, 
whereas succeeded in shock revival for 2D and 3D models. 
First we summarize shock evolution driven by neutrino heating (\ref{sec:3d-shock}), 
then examine neutrino properties as well as PNS convection which affects neutrino emission (\ref{sec:3d-neu}). 
We also estimate some observable values 
adopting the current and future neutrino and gravitational wave (GW) detectors (\ref{sec:obs}).

\subsection{Shock evolution} \label{sec:3d-shock}
\begin{figure}
    \centering
        \includegraphics[width=1\linewidth]{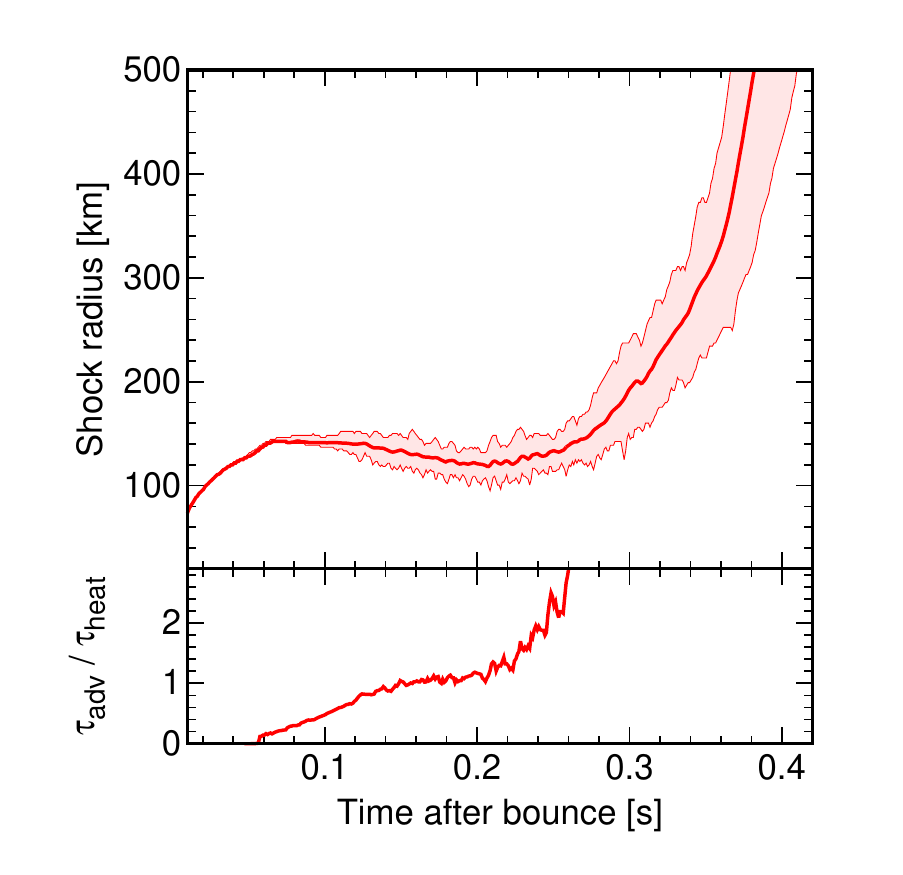}
    \caption{Top panel: The early time evolution of the shock radius for the 3D model. 
    The coloured-in region indicates the range of the shock location from minimum to maximum. 
    The shock once stalls at $r \sim 150$\,km, 
    gradually retreats, 
    then turns to expand at 220\,ms after bounce. 
    Shown in the bottom panel is a ratio of advection timescale to neutrino heating timescale. 
    The timescale ratio clearly exceeds unity just prior to the shock revival, which suggests that the shock revival is driven by neutrino heating.}
    \label{fig:rsh-3d}
\end{figure}

\begin{figure}
    \centering
        \includegraphics[width=1\linewidth]{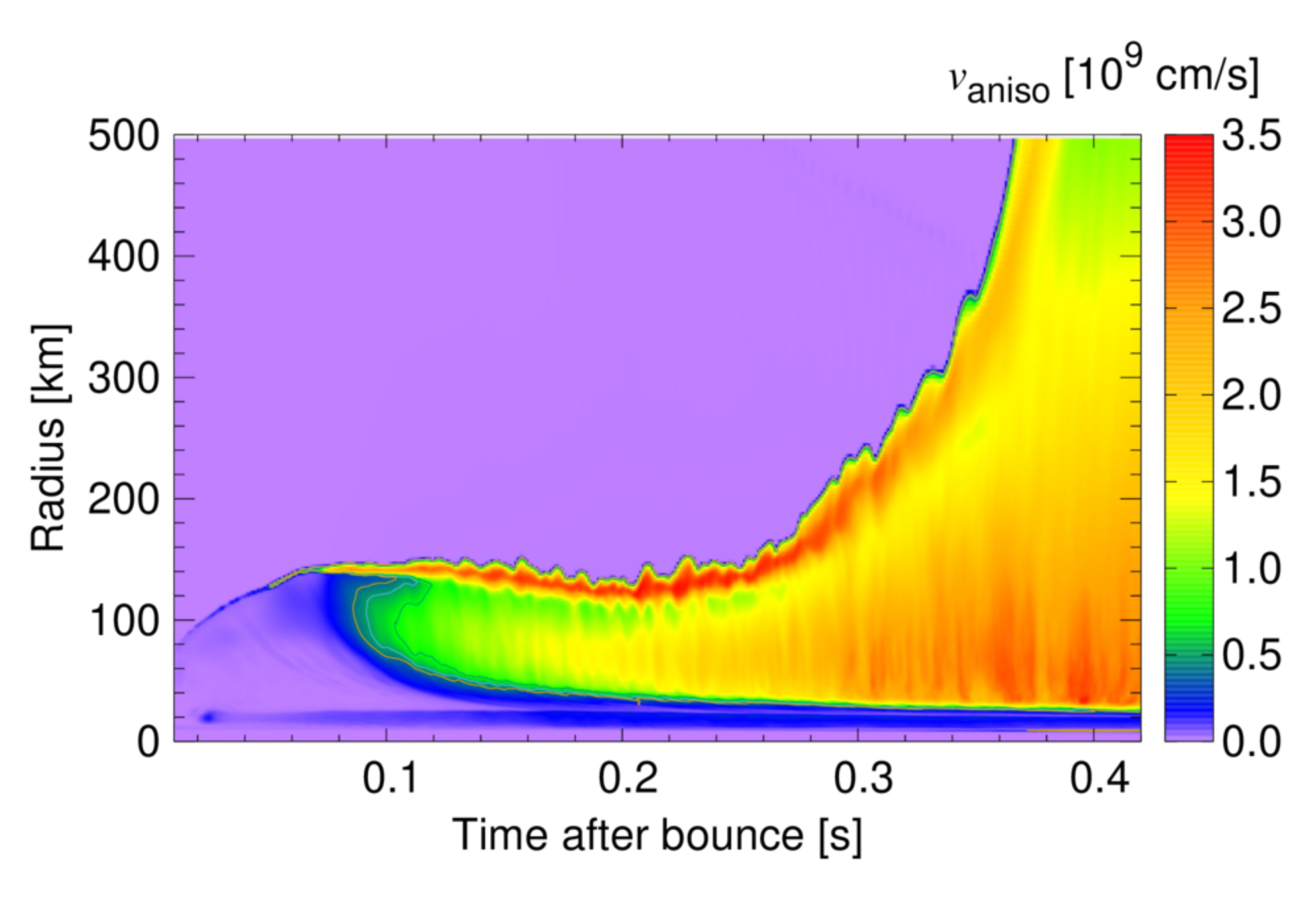}
    \caption{A space-time diagram of the anisotropic velocity. 
    The yellowish region with high anisotropic velocity appears in the gain region at $t_{\rm pb} \sim 150$\,ms, then the turbulent region spreads and pushes the stalled shock outward.}
    \label{fig:v-aniso}
\end{figure}

Figures \ref{fig:rsh-3d} and \ref{fig:v-aniso} show early evolution of shock radius 
and some relevant values representing the condition of neutrino-heated matter behind the shock.
The shock of our 3D model once stalls at $r \sim 150$\,km, 
gradually shrinks down to $\sim 120$\,km, 
then turns to expand at $\sim 220$\,ms after bounce 
(top panel of Figure~\ref{fig:rsh-3d}). 
The corresponding 1D simulation does not explode, whereas it is not shown in Figure~\ref{fig:rsh-3d}. 
To assess the mechanism of the shock revival 
we estimate and compare two timescales, the advection time, 
$\tau_{\rm adv} = \int \frac{dr}{|v_r|}$, 
and the heating time,
$\tau_{\rm heat} = \frac{\int e_{\rm bind} dV}{\int \dot{q} dV}$, 
where $v_r$ is the angle-averaged radial velocity, 
$e_{\rm bind}$ is the binding energy of the matter, 
$\dot{q}$ is the local heating rate by neutrino, 
and all of them are integrated over the gain region. 
Here the gain region is the region 
where the neutrino heating overcomes neutrino cooling, 
located behind the shock and extended to typically 50--60\,km from the centre.
The ratio of the advection to heating timescales are shown in the bottom panel of Figure~\ref{fig:rsh-3d}. 
Preceding to the shock revival, the time ratio clearly exceeds one at $\sim 200$\,ms after bounce, 
which means that the matter behind the shock could become unbound by neutrino heating before it passes through the gain region.

To visualize the motion of matter caused by neutrino heating, 
we present anisotropic velocity profile in Figure~\ref{fig:v-aniso}. 
Here we define the anisotropic velocity, $v_{\rm aniso}$, following \citet{takiwaki12} as
\begin{equation}
v_{\rm aniso} 
= \sqrt{\langle \rho \left((v_r- \langle v_r \rangle)^2 
        + v_\theta^2 + v_\phi^2 \right) \rangle
        / \langle \rho \rangle},
\end{equation}
where the symbol $\langle x \rangle$ denotes angle average of $x$. 
We see that anisotropic motion in the gain region appears 
around $r \sim 80$\,km at $\sim 150$\,ms after bounce. 
The turbulent motion develops and extends to the shock front at $\sim 220$\,ms after bounce, then the shock expansion accelerates and never retreats. 

\begin{figure*}
  \begin{center}
  \begin{tabular}{ccc}
  \includegraphics[width=0.3\linewidth]{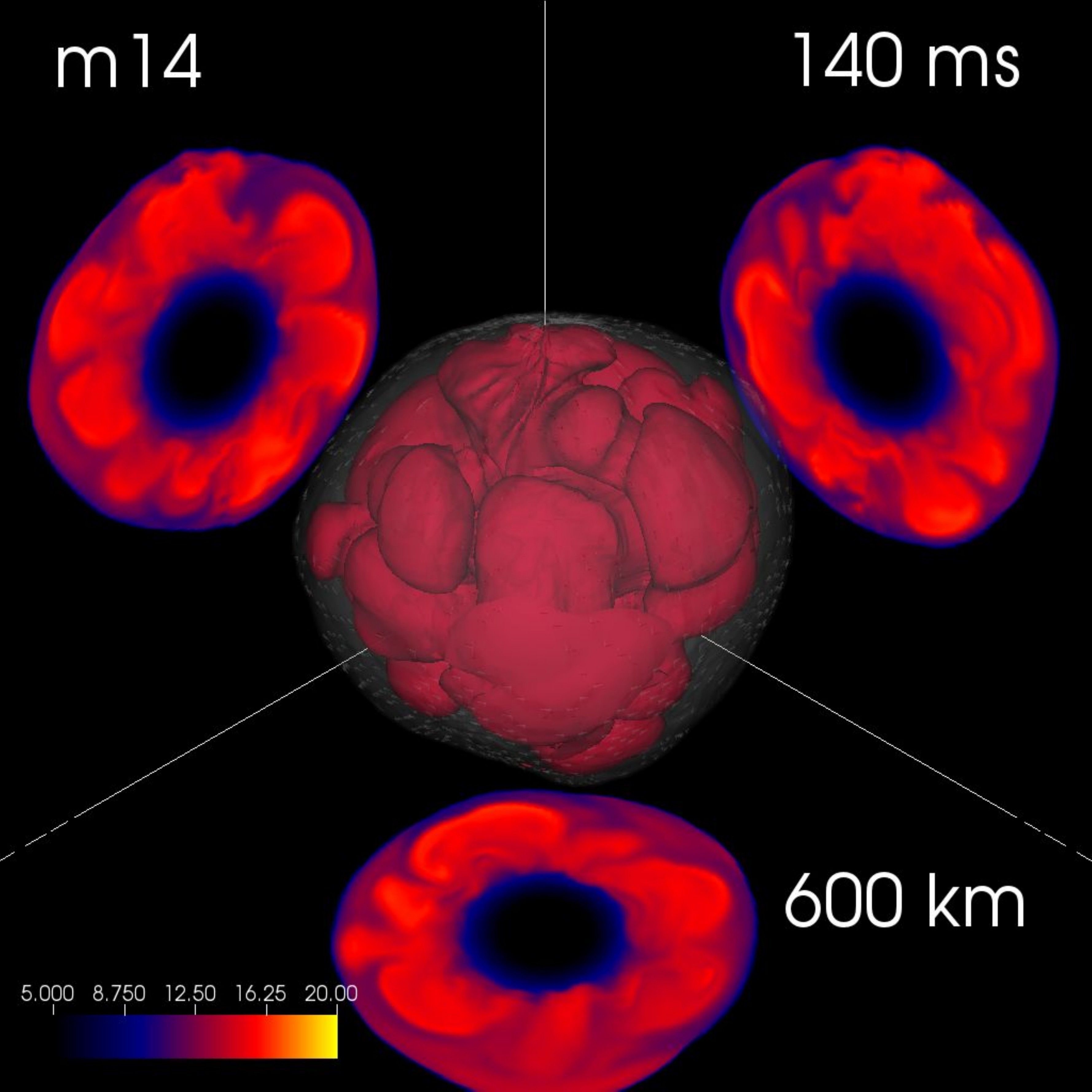} &
  \includegraphics[width=0.3\linewidth]{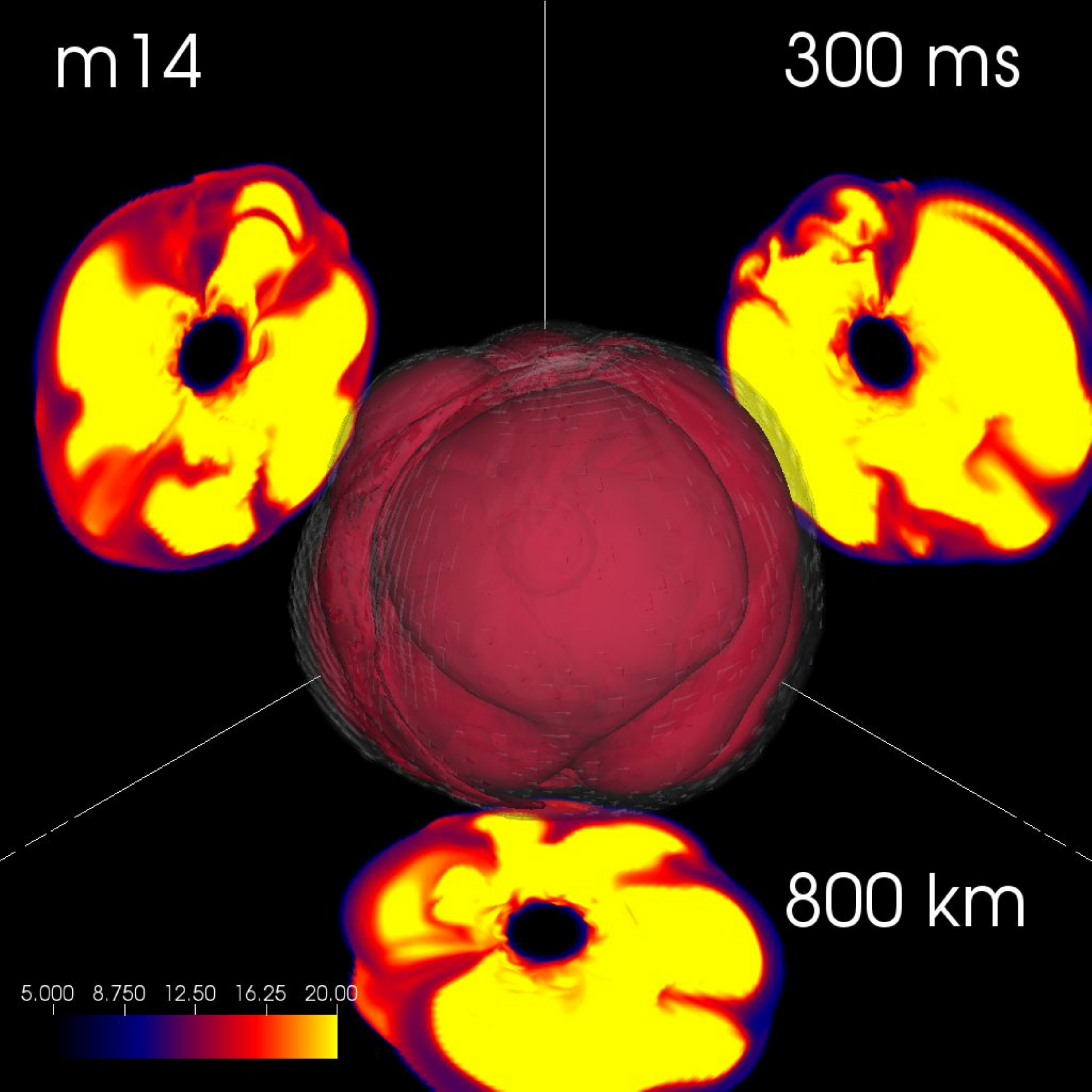} &
  \includegraphics[width=0.3\linewidth]{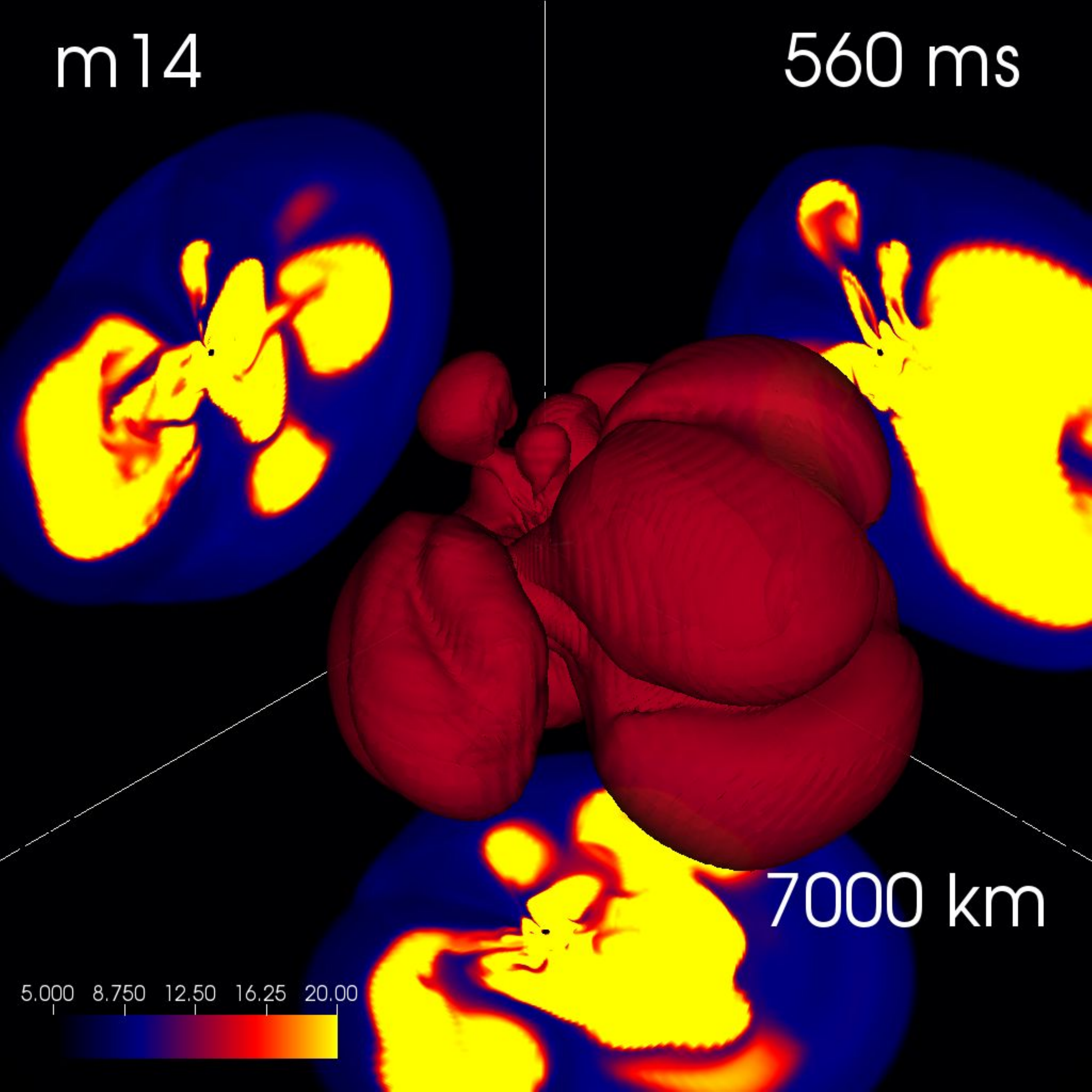} \\
  \end{tabular}
  \end{center}
  \caption{Time sequence of the isosurface of entropy, showing the morphology of the hot bubbles within the inner core of the exploding SN 1987A progenitor. 
  Shown are three snapshots at 140\,ms (left panel), 300\,ms (middle), and 560\,ms (right) after bounce. 
  The time and spatial scales (box size) are noted in each panel. 
  The contours on the cross-sections in the $x = 0$, $y = 0$, and $z = 0$ planes are projected on the walls at the back right, back left, and bottom of each panel, respectively. 
  Several high-entropy hot bubbles are visible, which push the shock outward without any dominant direction.}%
  \label{fig:snap-3d}
\end{figure*}

The 3D morphology of the shocked matter and the highly non-axisymmetric nature of the explosion are visualized in Figure~\ref{fig:snap-3d}. 
By $\sim 150$\,ms after bounce, small-scale convective motion develops behind the shock and it slightly deforms the shock structure (left panel). 
The continuous neutrino heating and the convective motion finally make the stalled shock turn to expand. 
The expanding shock deviates from spherical symmetry, but it has no specific direction in our 3D model. 

\subsection{Neutrino emission and PNS convection} \label{sec:3d-neu}
\begin{figure*}
  \begin{center}
  \begin{tabular}{cc}
    \includegraphics[width=0.45\linewidth]{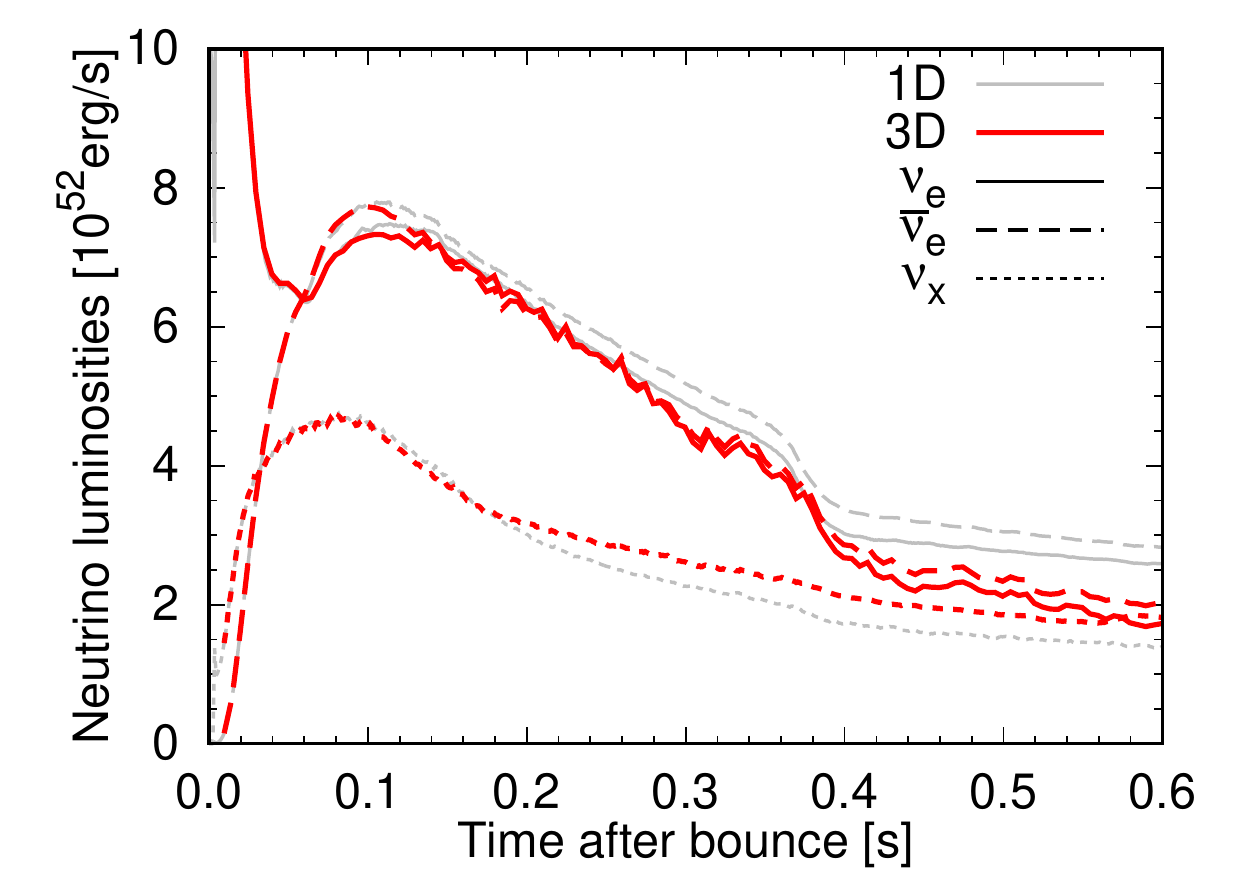} &
    \includegraphics[width=0.45\linewidth]{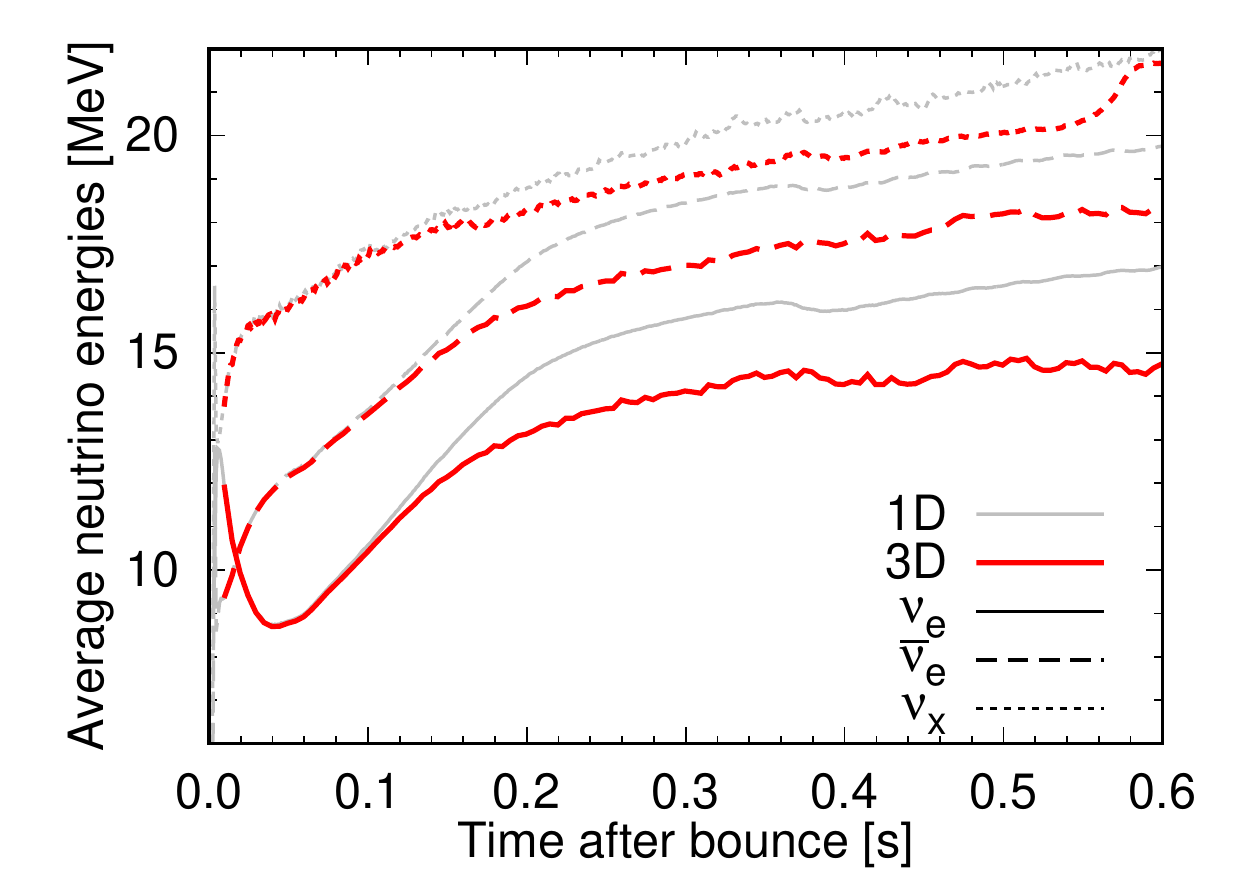} \\
  \end{tabular}
  \end{center}
  \caption{Neutrino luminosity (left panel) and average neutrino energy (right panel) as a function of time after bounce. 
  The luminosities and average energies for the 3D model (thick red lines) are compared with those for the 1D model (thin grey lines), 
  showing smaller values except $\nu_{\rm x}$ luminosity under the influence of the shock revival followed by a feeble accretion in the 3D model. 
  }%
  \label{fig:nu1d3d}
\end{figure*}

The behaviour of the matter behind the shock and the shock itself implies that neutrino heating plays a critical role in our 3D explosion model.
Figure~\ref{fig:nu1d3d} shows luminosity and average energy of 
electron, anti-electron, and heavy lepton neutrinos ($\nue$ , $\nub$, and $\nux$) 
of our 3D model (thick red lines) compared with that of 1D model (thin gray lines) for reference. 
For the $\nue$ and $\nub$ luminosity, both 1D and 3D models show sudden drop at $\sim 360$\,ms after bounce (left panel). 
Prior to this luminosity drop, the Si/O interface falls onto the central region. 
The mass accretion rate decreases, resulting in a decrease in the release rate of 
gravitational energy carried by the accreting matter. 
The decrease of the accretion also causes a shock expansion in the 3D model, 
and the accretion rate drops further. 
In contrast, the 1D model does not present shock revival 
and keeps relatively high neutrino luminosity compared with the 3D model.

On the other hand, 
The $\nux$ luminosity does not show such a sudden drop since $\nux$ mainly comes from deep inside the core. 
Moreover, the $\nux$ luminosity of the 3D model is higher than that of the 1D model. 
The deviation of the $\nux$ luminosity between 1D and 3D models becomes visible at $\sim 150$\,ms after bounce, which is roughly corresponding to the time when the anisotropic motion becomes outstanding in the 3D model (Figure~\ref{fig:v-aniso}).
The turbulent motion seen in Figure~\ref{fig:v-aniso}, however, first appears at $r \sim 80$\,km, 
which is out of neutrino spheres.
It implies that this turbulent motion in the 3D model does not play a significant role in the deviation of neutrino properties from the 1D model.



\begin{figure}
  \begin{center}
    \includegraphics[width=1\linewidth]{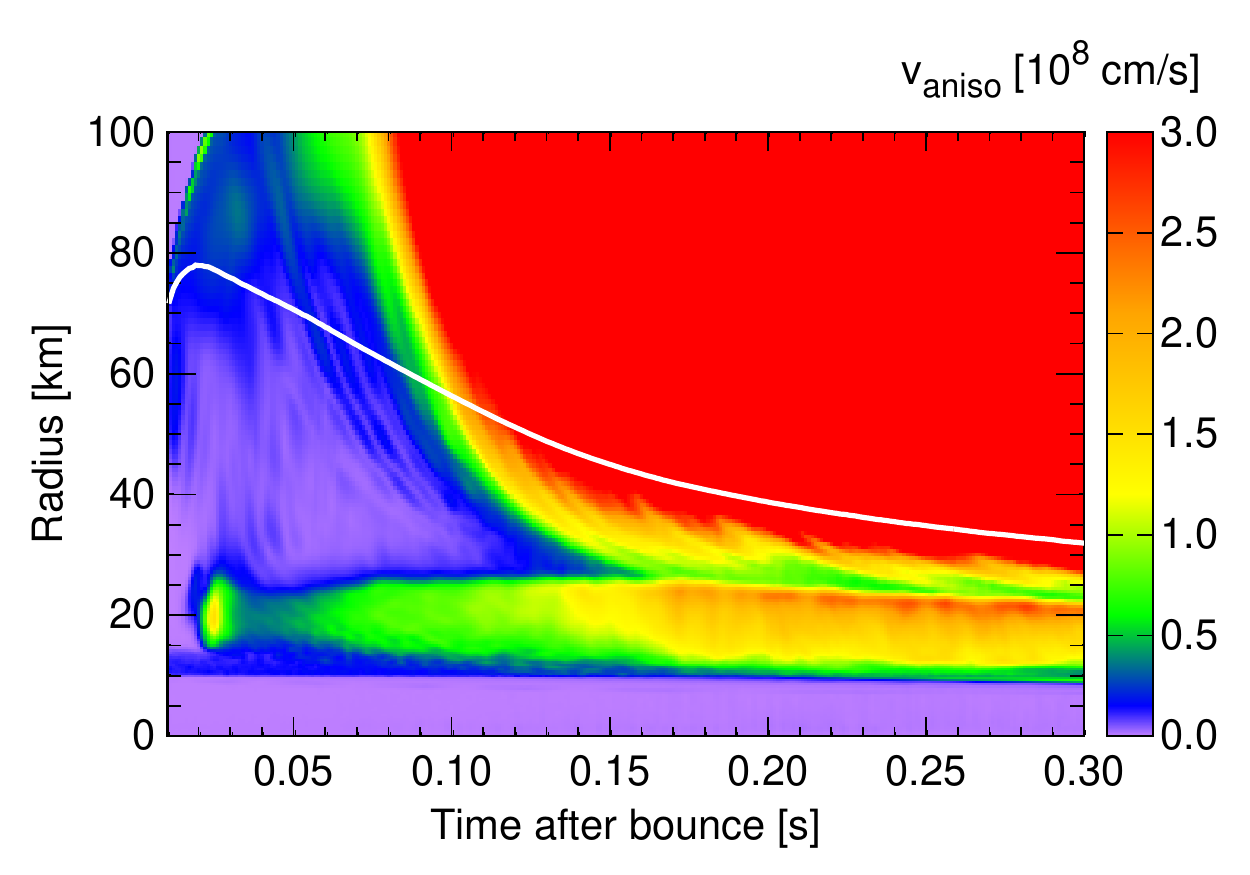}
  \end{center}
  \caption{A space-time diagram of the anisotropic velocity for the 3D model as in Figure~\ref{fig:v-aniso}, but within the inner 100\,km through 0.3 s after bounce. 
  The inner PNS convective region is visible around $r \sim 20$\,km. 
  The white line illustrates the mean PNS radius defined as the radius where mass density $\rho = 10^{11} \, {\rm g \,cm}^{-3}$.
  }%
  \label{fig:v-aniso-pns}
\end{figure}

Again, since $\nux$ mainly comes from deep inside the core, 
we provide a space-time diagram of the anisotropic velocity within the inner 100\,km 
in Figure~\ref{fig:v-aniso-pns} to assess hydrodynamic motions deep inside the core. 
Two convective regions, the exterior neutrino-driven convection as in Figure~\ref{fig:v-aniso} and the interior convective band around $r \sim 20$\,km, are visible in this zoom-in panel. 
The interior convective zone first emerges at $\sim 25$\,ms after bounce, 
then once becomes weaken and again active at $\sim 150$\,ms. 
This may cause effective transport of diffusive neutrinos inside the PNS and explain the increase of $\nux$ luminosity in the left panel of Figure~\ref{fig:nu1d3d} as discussed in \citet{buras06a} 
and \citet{nagakura20}. 
Furthermore, the interior convective motions hold up the PNS structure and the PNS radius of the 3D model becomes larger than that of the 1D model. 
This leads to a lift of neutrino sphere radius and a 1--2 MeV decrease of average neutrino energy of the 3D model compared to that of the 1D model 
as observed in the right panel of Figure~\ref{fig:nu1d3d} after $\sim 150$\,ms and later. 

\begin{figure}
  \begin{center}
      \includegraphics[width=0.95\linewidth]{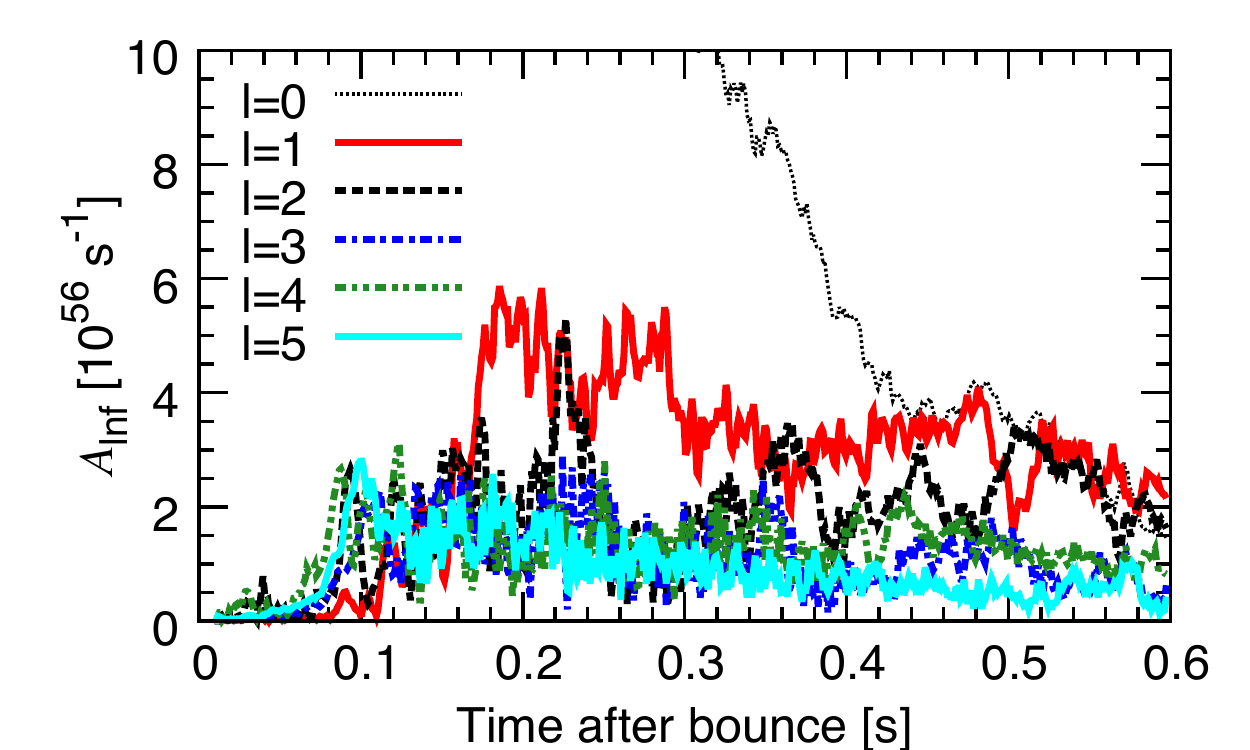}
  \end{center}
  \caption{Monopole and multipole moments of the electron lepton-number flux for 3D models, 
  evaluated at a radius of $r=500$\,km and transformed into the lab frame at infinity, 
  as functions of time after core bounce. 
  The multipole amplitudes $A_{\rm lnf}$ are defined as given in Equation (\ref{eq:lesa}), 
  while for $l=0$ the corresponding amplitude agrees with the total electron lepton-number flux. 
  }%
  \label{fig:lesa}
\end{figure}

\citet{tamborra14b} found in their 3D CCSN model that 
the lepton-number flux ($\nue$ minus $\nub$) emerges predominantly in one hemisphere, 
termed ``Lepton-number Emission Self-sustained Asymmetry (LESA)''.
We evaluate signatures of LESA in our 3D model.
Following \citet{glas19b}, the moments of the lepton-number flux are evaluated as
\begin{equation}\label{eq:lesa}
    A_{\rm lnf}(r,l) = \sqrt{\sum_{m=-l}^{l}[c_l^m(r)]^2},
\end{equation}
where the multipole coefficients $c_l^m(r)$ at a given radius $r$ are calculated by integrating the flux over the whole spherical surface
\begin{equation}
    c_l^m(r) = \sqrt{\frac{4 \pi}{2l+1}} \int r^2 F_{\rm lnf}^{\rm lab} Y_l^m d\Omega ,
\end{equation}
using the radial number flux of neutrino $F_{\rm lnf}^{\rm lab}$ in the lab frame.
The time evolution of the multipole moments $A_{\rm lnf}$ of the electron lepton-number flux 
is shown in Figure~\ref{fig:lesa}.
High order modes with $l=4$ and $l=5$ are excited first at $\sim 60$\,ms after bounce. 
The emergence of these asymmetric neutrino fluxes roughly coincides with the time 
when the convective motion inside the PNS sets in (Figure~\ref{fig:v-aniso-pns}).
A dominant dipole ($l=1$) begins to appear at $\sim 170$\,ms after bounce 
and the dipole and quadrupole ($l=2$) modes have dominant amplitudes until the end of our simulation. 
Such a time-evolving mode shift is consistent with \citet{glas19b} 
who examine the effects of LESA with different progenitors, neutrino transport treatment, and grid resolutions. 

\subsection{Multi-messenger observables} \label{sec:obs}

\begin{figure*}
  \begin{center}
  \begin{tabular}{cc}
    \includegraphics[width=0.45\linewidth]{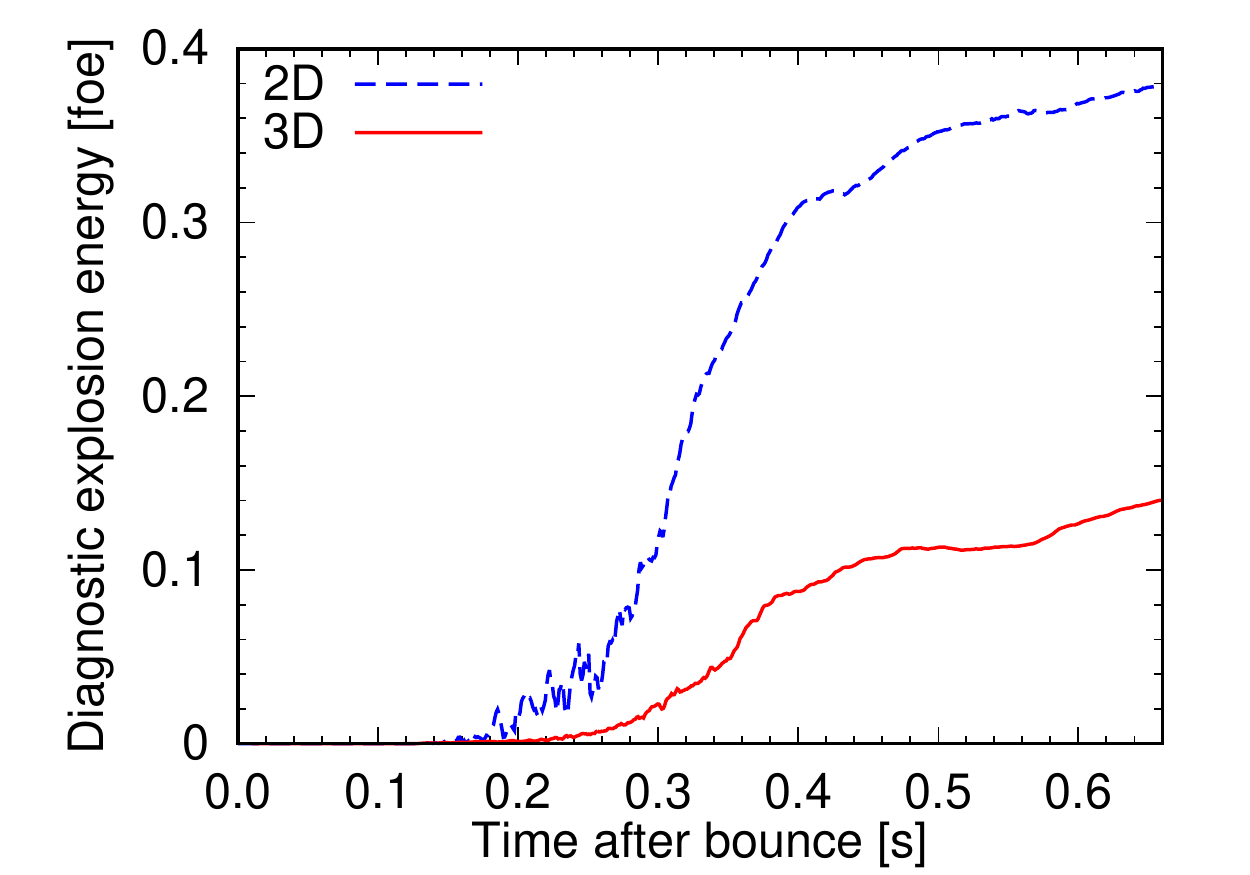}   &
    \includegraphics[width=0.45\linewidth]{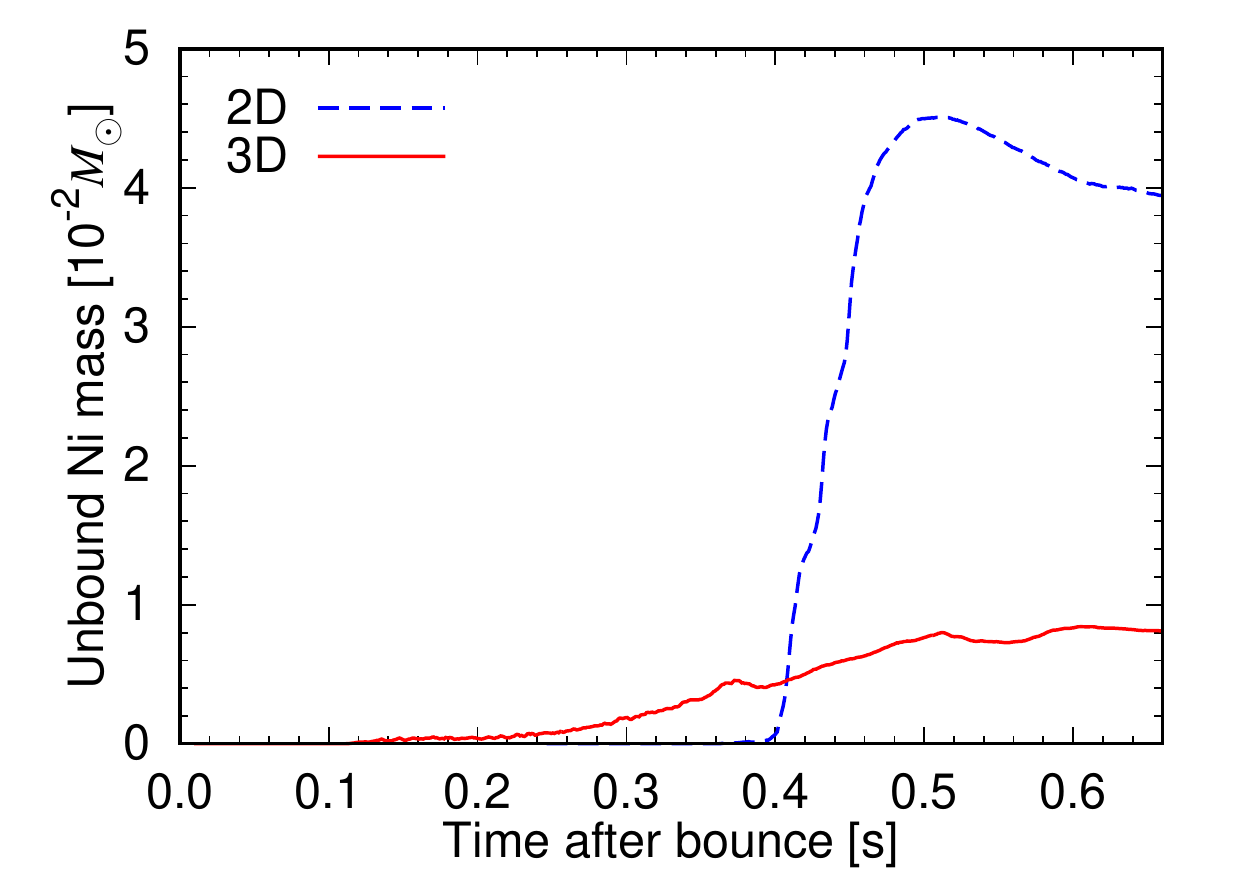}\\
  \end{tabular}
  \end{center}
  \caption{Time evolution of diagnostic explosion energy (left panel)
  and unbound Ni mass (right panel) of 2D and 3D models.}%
  \label{fig:eexp_mni-2d3d}
\end{figure*}

At the end of our 3D simulation ($\sim 660$\,ms after bounce), 
the average and the maximum shock radius have reached at $\sim 3000$\,km and $\sim 4400$\,km.
The diagnostic explosion energy grows up to $\Ediag \sim 0.14 \times 10^{51}$\,erg by this time (Figure \ref{fig:eexp_mni-2d3d}). 
Here the diagnostic explosion energy is defined as the sum of internal, kinetic, and gravitational energy in unbound materials with positive radial velocity. 
The electromagnetic lightcurve of SN 1987A is different from that of typical type IIP supernovae, 
but it can be reproduced by an explosion of a BSG with explosion energy 
$\Eexp \sim 1.5 \times 10^{51}$\,erg \citep{jerkstrand20}, 
one order of magnitude larger than the diagnostic explosion energy of our 3D model. 
The feeble explosion results in a small mass of $^{56}$Ni yield. 
We obtain a rough estimate of $^{56}$Ni mass, $\Mni$, in the unbound materials less than $0.01\,\Msun$ 
by means of a simple 13 $\alpha$ network calculation, 
whereas the the value estimated from observations is $\Mni \sim 0.07\,\Msun$ \citep{bouchet91,mccray93,seitenzahl14}.
Note that \citet{sawada19} and \citet{sawada21} argue that the timescale of the shock revival affects the $^{56}$Ni yield. Their analysis indicates that the shock revival should occur earlier to obtain more yields.

The time evolution of the explosion energy and $^{56}$Ni are compared to the results in 2D simulation in Figure~\ref{fig:eexp_mni-2d3d}. 
The 2D model assuming axi-symmetry presents a more energetic explosion with $\Ediag \sim 0.38 \times 10^{51}$\,erg and $\Mni \sim 0.04\,\Msun$. 
The reason of the weak explosion in our 3D model will be discussed in \S \ref{sec:summary}.

\subsubsection{Neutrino}
\begin{figure}
  \begin{center}
    \includegraphics[width=1\linewidth]{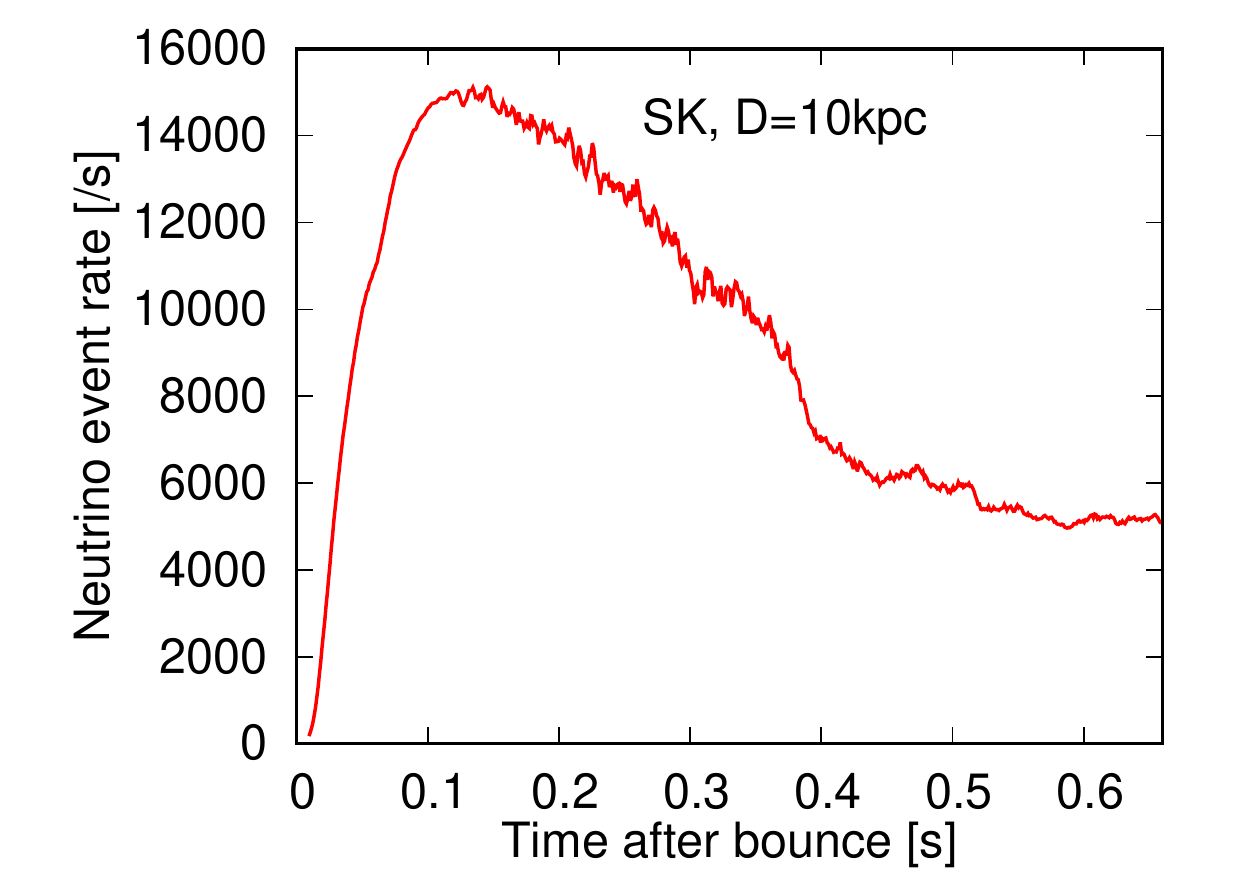}
  \end{center}
  \caption{Expected neutrino events at Super-Kamiokande (SK) in 1 second time bins. 
  Here we assume SK detector with an effective volume of 32\,kton 
  and the distance to the CCSN event of 10\,kpc.}%
  \label{fig:nuobs}
\end{figure}

Our 3D model is not capable of direct comparison with SN 1987A observations. 
Below in this subsection, we assume that a supernova modelled by our 3D simulation occurs at the Galactic centre at the distance of 10 kpc 
and evaluate the multi-messenger signals 
like neutrino and gravitational wave from the CCSN. 
Figure~\ref{fig:nuobs} shows the neutrino detection event rate per 1\,s bin expected 
by the Super-Kamiokande (SK) detector for the 10\,kpc CCSN event. 
Here we only take account of inverse-beta events and do not consider MSW mixing (e.g., \citet{kotake06} for a review) and collective neutrino oscillations 
\citep[see, e.g.,][for collective references therein]{Mirizzi16,horiuchi18_rev,tamborra19,tamborra21,sajad19,chakra20,cherry20,sasaki20,sasaki22ax,kato20,kato21,richers21,zaizen21,nagakura21c,nagakura21d,sigl22}. 
The high-statistics event curve will bring us plenty of information.
The sudden drop at 0.35--0.4 s after bounce corresponds to the time of shock revival 
when the stalled shock turns to expand and neutrino luminosity from accreting matter decreases. 
Another revelation is a precise estimate of the bounce time, which is critical to reducing the background noise for gravitational wave detection \citep[see][for the discussion of multimessenger signals from a Galactic CCSN]{nakamura16}. 

To obtain the whole neutrino light curve and spectrum from the supernova, we have to perform long term simulations, whose simulation time should be longer than $10\,{\rm s}$ \citep[e.g.,][]{suwa19,nakazato22}, which is beyond the scope of this paper.
The relation between the character of PNS (e.g., mass and radius) and the neutrino luminosity is an important subject to investigate \citep{nakazato19,nagakura21ax}.
For this purpose an analytic solution of the light curve would be useful \citep{suwa21}. 

\subsubsection{Gravitational wave}
\begin{figure*}
  \begin{center}
  \begin{tabular}{cc}
    \includegraphics[width=0.45\linewidth]{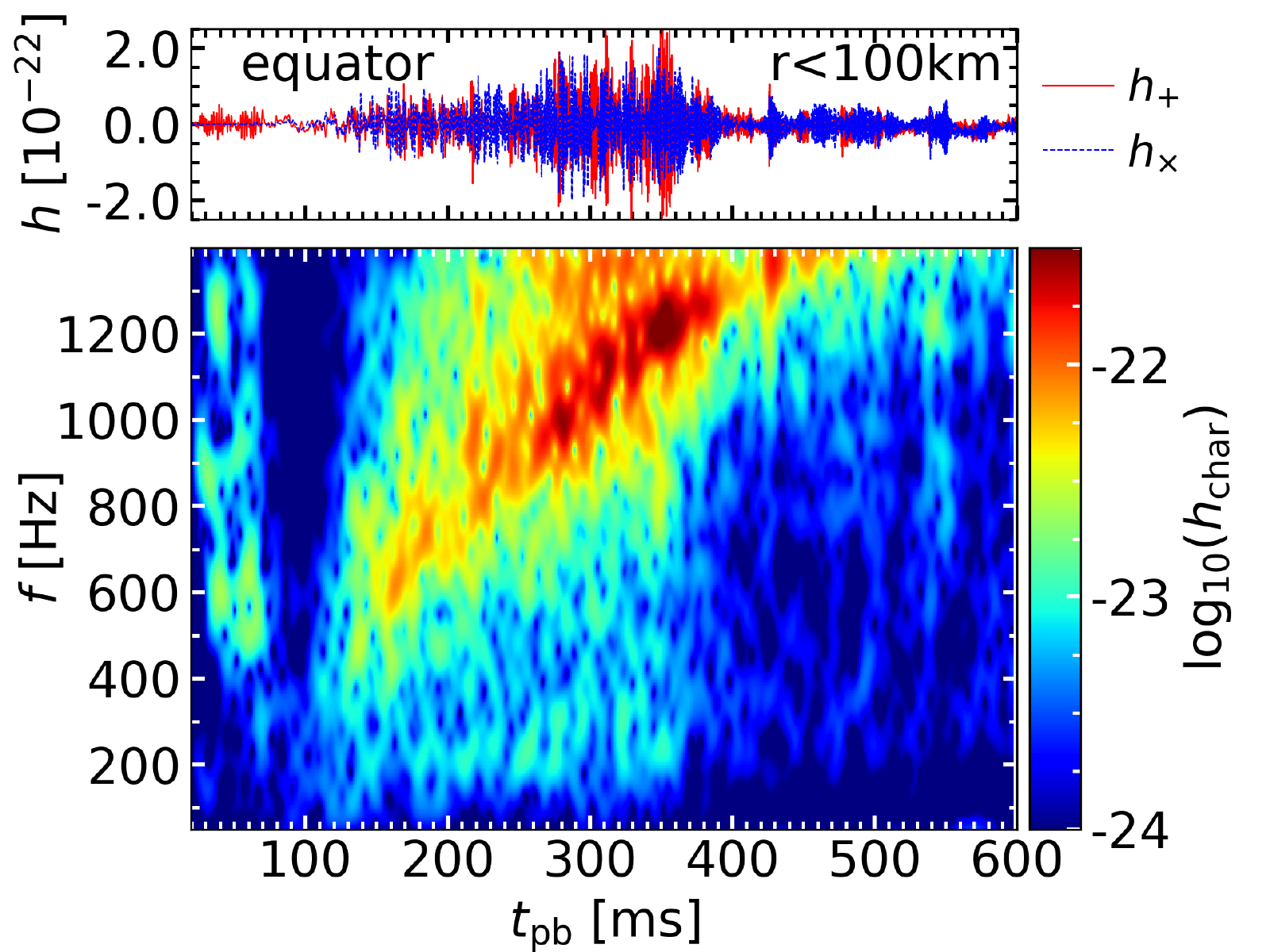} &
    \includegraphics[width=0.45\linewidth]{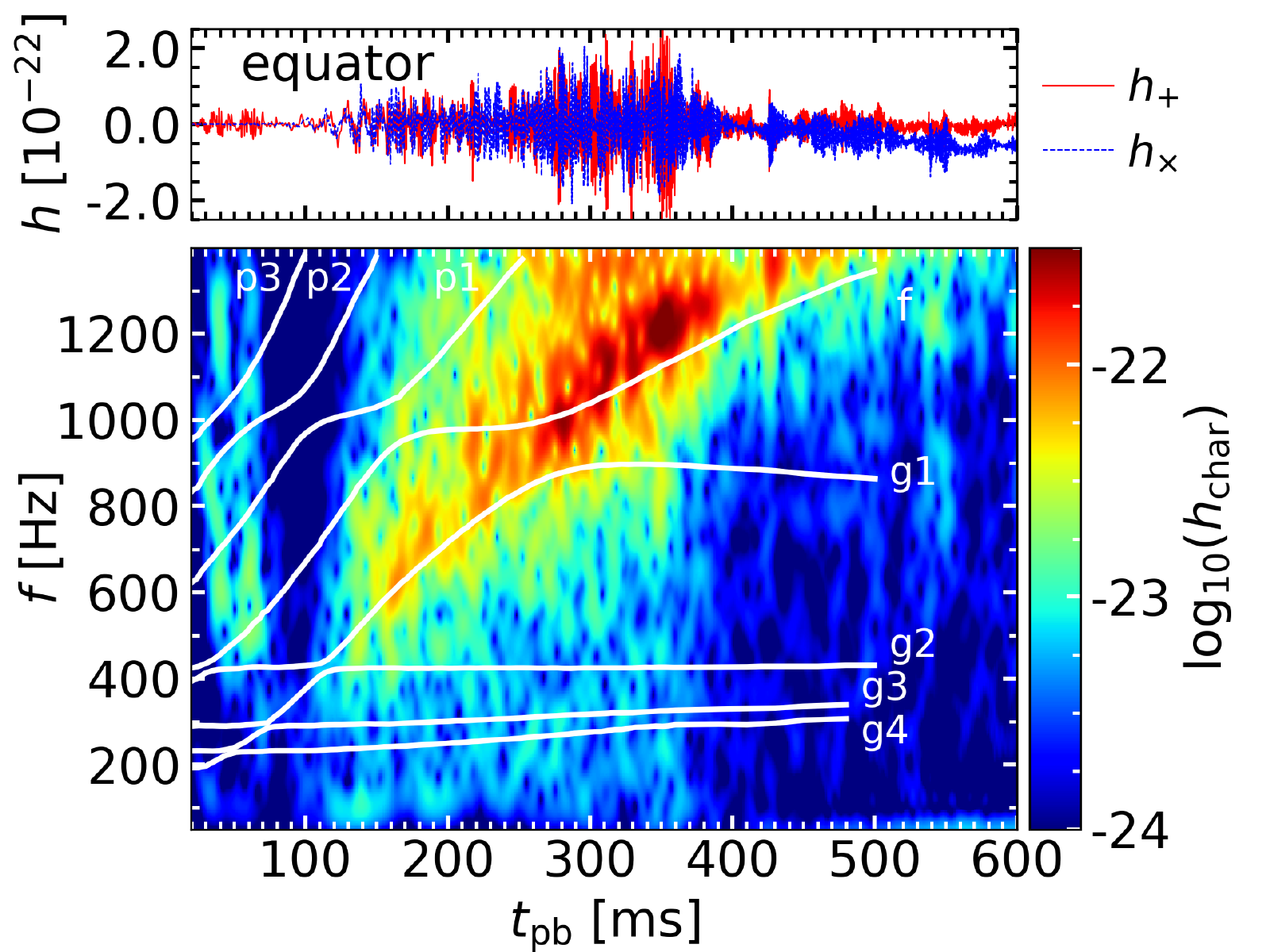}
  \end{tabular}
  \end{center}
  \caption{GW waveform and spectrogram of our 3D model viewed from the equator ($x$-axis) direction.
  Left panel: the input waveform in the top window and the Fourier-transformed spectrogram in the bottom window. GW source is limited within $r < 100$ km. 
  Right panel: same as the left panel but for the total GW signal. Overplotted are some characteristic gravitational wave frequencies calculated with GREAT.}
  \label{fig:gw-spectrogram-eq}
\end{figure*}

\begin{figure*}
  \begin{center}
  \begin{tabular}{cc}
    \includegraphics[width=0.45\linewidth]{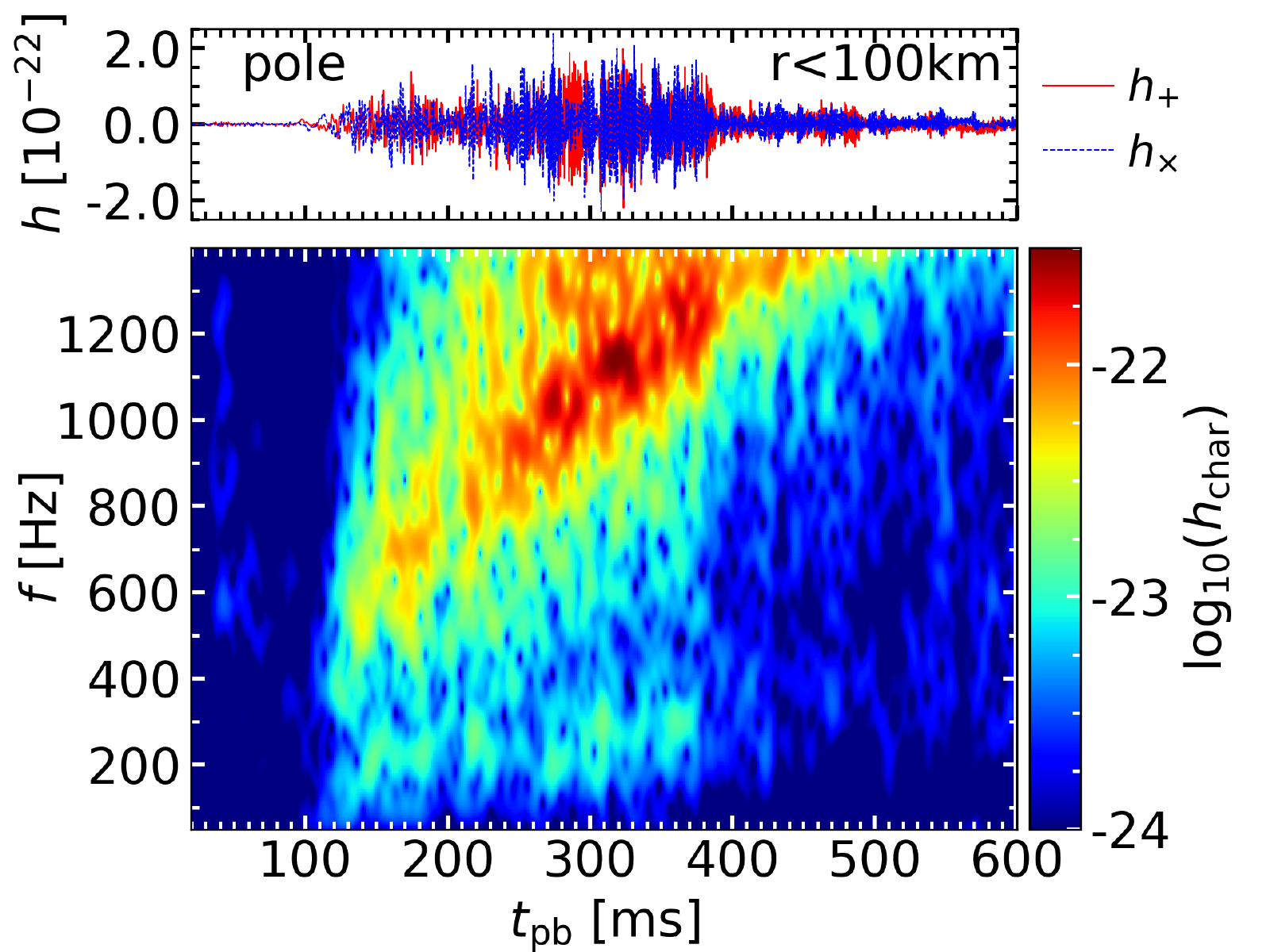} &
    \includegraphics[width=0.45\linewidth]{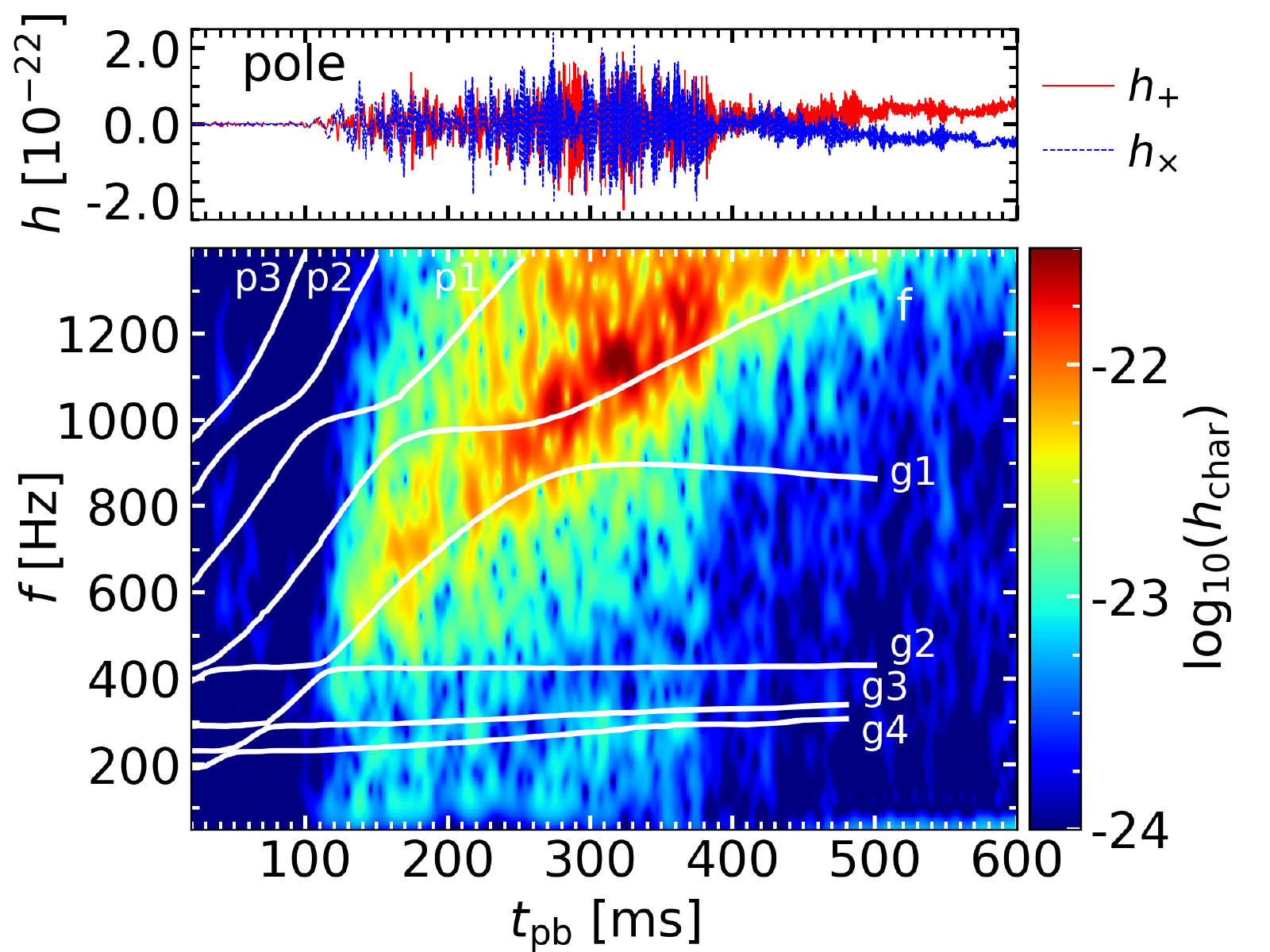}
  \end{tabular}
  \end{center}
  \caption{Same as Figure~\ref{fig:gw-spectrogram-eq} but viewed from the polar ($z$-axis) direction.
    }
    \label{fig:gw-spectrogram-pl}
\end{figure*}

A gravitational-wave (GW) signal from CCSN is very weak compared to that from compact merger event.
Once the neutrino detection event tells us the time of core bounce, however, 
a small time window for analysis is specified and background noise is greatly suppressed 
so that we expect gravitational wave detection from a Galactic core-collapse event
\citep{nakamura16}. 
This detection is expected to revolutionize our CCSN theory
\citep[see, e.g.,][for collective references therein]{kotake13,kuroda16,andresen19,radice19,ernazar2020,BMuller20,mezzacappa20, Andersen21}
because the GW signals deliver the live information to us, which imprints the non-spherical hydrodynamics evolution up to the onset of explosion.
Figures \ref{fig:gw-spectrogram-eq} and \ref{fig:gw-spectrogram-pl} 
show GW waveform of plus and cross modes ($h_+$ and $h_\times$, top windows) 
and a colour map of frequency $f$ components (spectrogram, bottom windows) as a function of time after bounce. 

In Figure~\ref{fig:gw-spectrogram-eq} we assume that an observer is along the $x$-axis of the simulation coordinates. 
One can see a typical time series of CCSN-GW characteristics 
both in plus and cross modes: 
first (0 to $\sim 50$\,ms), a GW signal from prompt convection induced by a negative entropy gradient after the stalling shock appears 
in the wide range of frequency, 
followed by a relatively silent phase lasting from $\sim 50$ ms to $\sim 150$\,ms, 
and then nonlinear fluid motions behind the shock excite strong high-frequency signals 
(starting from $f \sim 500$~Hz at 150 ms and growing up to $f > 1000$~Hz at $\sim 400$\,ms), 
which is attenuated as the shock expands and the nonlinear motions behind the shock are diluted ($t_{\rm pb}>400$~ms).
Another component can be seen at around 100--300\,Hz after $\sim 150$\,ms. 
\citet{kuroda16} pointed out that the low frequency GW excess is correlated to the SASI in a similar frequency range \footnote{More recently the GW frequency is identified as a doubling of the SASI frequency (e.g., \citet{andresen19,Shibagaki20,shibagaki21,takiwaki21}).}. However, it should be noted that our model does not present SASI-like motion. 
This low-frequency signal in our model seems to correspond to a turn over timescale of the inner PNS convective motions (Figure~\ref{fig:v-aniso-pns}) 
which emerges at $\sim 150$\,ms in the radius around $10^6$ cm with 1--3 $\times 10^8 \, {\rm cm \, s}^{-1}$
(see \citealt{mezzacappa20} and discussion in \citealt{Andersen21}, and also \citet{nagakura21a,nagakura21b} for the neutrino signature). 

In the left panels of Figures \ref{fig:gw-spectrogram-eq} and \ref{fig:gw-spectrogram-pl}, showing the GW signal from the central ($< 100$~km) matter, the amplitude is close to zero at the final phase.
On the other hand, the amplitude of the total signal in the right panels deviates from zero. 
This is caused by the non-spherical motions of matter behind the shock.

A view from the polar ($z$-axis) direction (Figure \ref{fig:gw-spectrogram-pl}) shows a similar time-evolution of GW waveform and spectrogram except for the prompt convection signal.
We cannot unambiguously
 identify the reason of 
 very weak prompt convection signal in the polar view. This may 
 stem from a stochastic nature of prompt convection, the growth of which could happen to 
  be direction-dependent.

We analyse a time-frequency behaviour of perturbative fluid motions  
by means of an open-source General Relativistic Eigen-mode Analysis Tool \citep[GREAT,][]{torres-forne19}
with changing the boundary condition \citep{Sotani20b}. 
We find some characteristic modes and overplot them on the GW spectrogram 
in the right panels of Figures \ref{fig:gw-spectrogram-eq} and \ref{fig:gw-spectrogram-pl}. 
The most dominant frequency component appears at $\sim 150$\,ms after bounce with $\sim 500$\,Hz 
and evolves to more than 1000\,Hz. 
This can be fitted by the $g1$ mode at first and shift to the $f$ mode at $\sim 250$\,ms after bounce \citep{sotani17,Morozova18,Sotani20c}. 
Relatively weak low-frequency (100--300\,Hz) components after $\sim 150$\,ms are fitted by the $g3$ and $g4$ modes.

\begin{figure*}
  \begin{center}
  \begin{tabular}{cc}
    \includegraphics[width=0.45\linewidth]{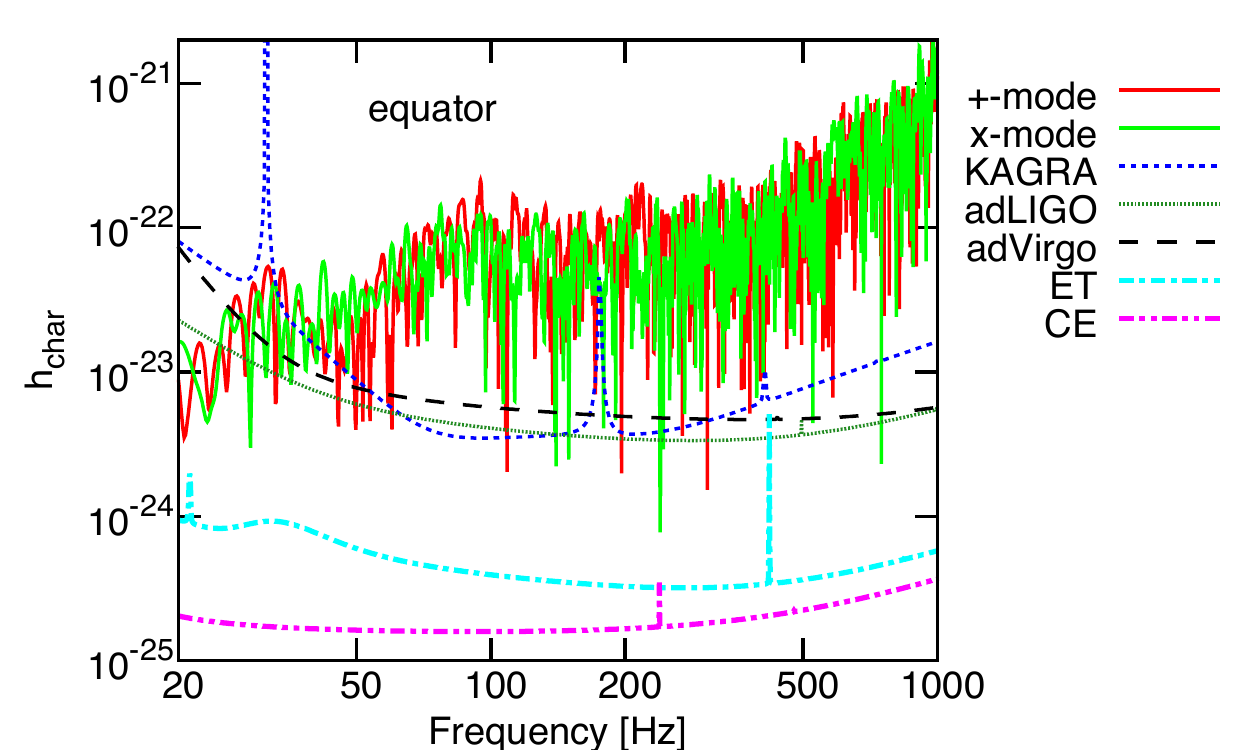}       &
    \includegraphics[width=0.45\linewidth]{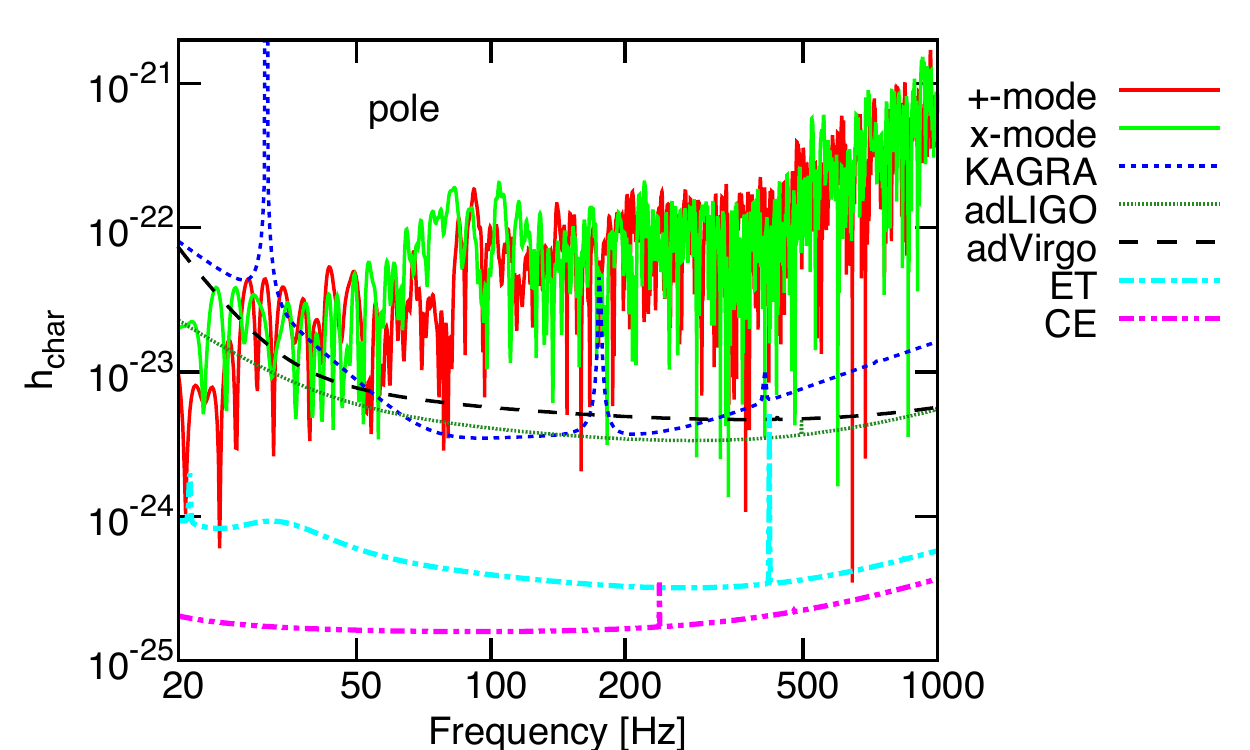} 
  \end{tabular}
  \end{center}
  \caption{
  The gravitational wave spectra of our 3D model (solid lines) are compared with some detection sensitivity curves of current and future GW detectors. Shown are two cases that observers stand along the $x$-axis (left panel) and along the $z$-axis (right panel).}%
  \label{fig:gw-spectra}
\end{figure*}

Figure~\ref{fig:gw-spectra} presents GW spectra of our supernova model as a function of GW frequency $f$, 
compared with the detection sensitivity curves of some current and future GW detectors. 
Here we again assume the distance to the supernova to be 10\,kpc. 
The detectors currently in operation (KAGRA, advanced LIGO, and advanced Virgo) 
have sensitivity to the GW signal for $f > 100$\,Hz with S/N ratio $>10$ 
and can be available for the analysis of the dominant high-frequency GW signals.
The design sensitivity of Einstein Telescope and Cosmic Explorer 
is good enough for a wide range of frequencies covering $f < 100$\,Hz. 
Although such next-generation detectors or very nearby supernova events are desirable, 
the possible PNS convection signal ($\sim 200$~Hz) could be detectable by the current detectors.

\subsubsection{Neutron star kick and spin}
\begin{figure*}
  \begin{center}
  \begin{tabular}{cc}
    \includegraphics[width=0.47\linewidth]{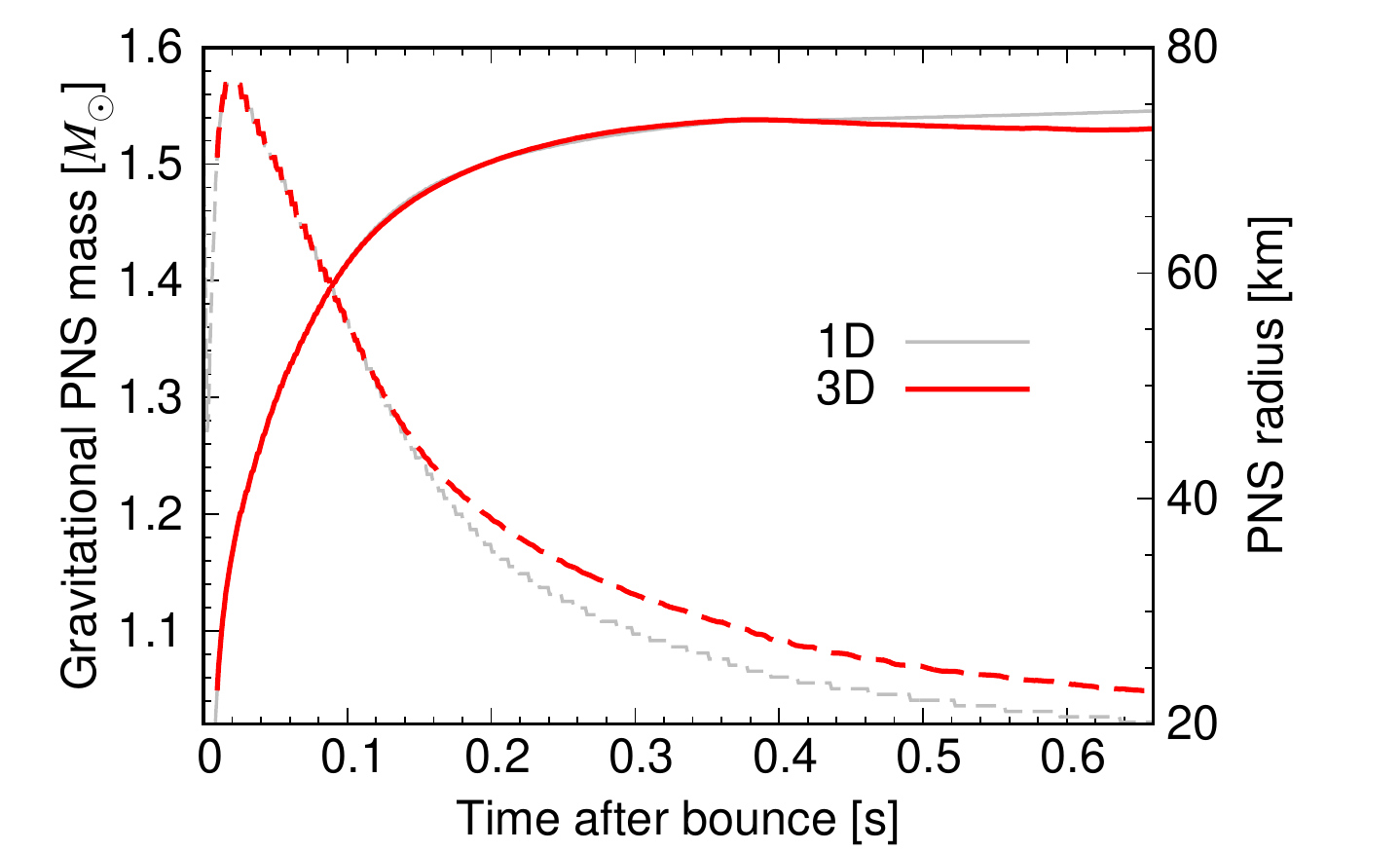} &
    \includegraphics[width=0.43\linewidth]{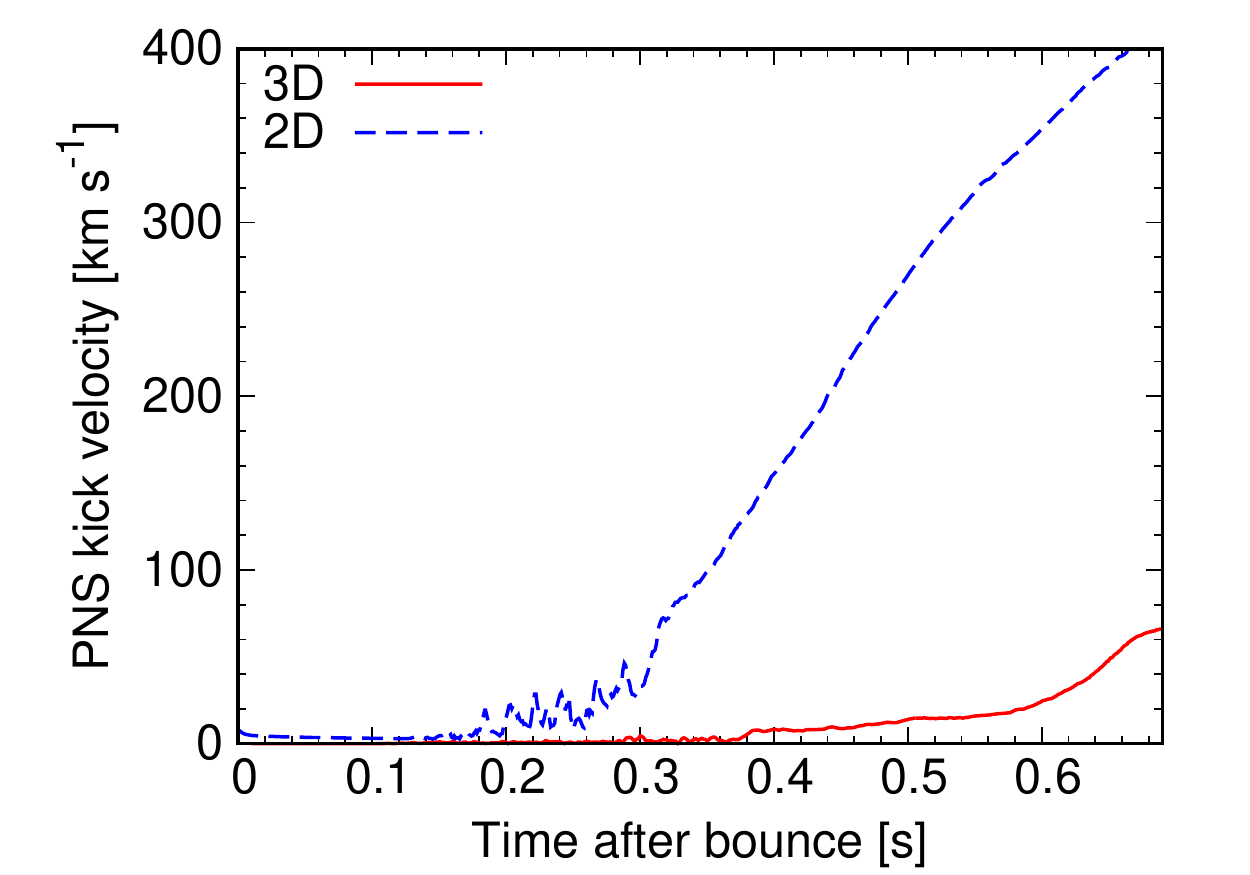}
  \end{tabular}
  \end{center}
  \caption{Left: The PNS mass (solid lines) and radius (dashed lines) 
  for the 1D (thin grey) and 3D (thick red) simulations as a function of time after bounce.
  The PNS surface is defined at a fiducial density of $\rho = 10^{11} \, {\rm g \, cm}^{-3}$.
  The larger PNS mass in 1D than in 3D is due to the long-lasting accretion in 1D. 
  At late times, the PNS radius for the 3D simulation is larger than the 1D case. This is presumably caused by PNS convection in the 3D model, as visible in Figure~\ref{fig:v-aniso-pns}. 
  The larger PNS radius in the 3D model, as well as neutrino sphere radius, result in lower neutrino energy (Figure~\ref{fig:nu1d3d}).
  Right: The PNS recoil velocity of the 3D model (solid red line), compared with that of the 2D model (dashed blue line). 
  The weak explosion with roughly spherical distribution of ejecta in 3D results in small recoil velocity ($\sim 70 \, {\rm km \, s}^{-1}$ at $t_{\rm pb} = 660$\,ms), 
  whereas axi-symmetric 2D model exhibits much larger kick velocity ($> 400\, {\rm km \, s}^{-1}$).
  }
  \label{fig:pns-kick}
\end{figure*}

A proper motion of the neutron star left behind the explosion is also meaningful. 
When a CCSN explosion is initiated in a non-spherical manner, 
the neutron star could be ``kicked'' by asymmetric hydrodynamic motions and mass ejection or by anisotropic neutrino emission from the PNS. 
Following the hydrodynamic kick scenario, we estimate the NS kick velocity as in \citet{nakamura19} (Figure~\ref{fig:pns-kick}). 
In this scenario, the kick velocity is determined by the net momentum of ejecta which is tightly connected to the explosion energy, asymmetricity of the ejecta motion, and the PNS mass. 
\citet{nakamura19} demonstrated long-term 2D CCSN simulations and concluded that the total momentum of 
the ejecta, or the explosion energy, is a predominant factor determining the strength of hydrodynamic kick, 
and the ejecta asymmetry plays a secondary role. 
Our 3D model has small explosion energy and roughly spherical mass distribution at least in the beginning of shock expansion, resulting in small kick velocity ($\sim 70 \, {\rm km \, s}^{-1}$ at $t_{\rm pb} = 660$\,ms, red solid line in the right panel of Figure~\ref{fig:pns-kick}). 
Explosion models associated with jet-like structures have been studied to make the fast-moving heavy elements and the observed luminosity in SN~1987A \citep{nagataki00} or to reproduce some morphological features of the remnant \citep{bear18b}. 
Our axi-symmetric explosion of the 2D model does not hold a collimated jet, but it 
presents more energetic explosion and highly asymmetric mass distribution than the 3D model, leading to higher kick velocity ($> 400\, {\rm km \, s}^{-1}$, dashed blue line). 

Optical, X-ray, and gamma-ray light curves of SN~1987A and broad emission lines of heavy elements have strongly suggested large-scale mixing during the explosion \citep{kumagai88,kumagai89,hachisu90}. 
Morphology and the properties extracted from photometric and spectral observations of SN 1987A 
have revealed that the distribution of the matter ejected from SN 1987A is completely non-spherical \citep{wang02,larsson13}. 
Moreover, there is an intricate triple-ring structure around SN 1987A, reflecting non-spherical mass ejection in the evolution of the pre-SN.
Observations of emission lines in the nebula phase have found a fast-moving iron clump and low-velocity hydrogen mixed down into the innermost ejecta, 
which can be reproduced by matter mixing via Rayleigh-Taylor instabilities at the chemical composition interfaces in the ejected envelope \citep{wongwathanarat15,utrobin15,utrobin19,utrobin21,ono20},
although it can not be captured by our space and time limited simulations. 
The NS kick is less affected by these asymmetries
since they develop at locations far from the central core and the kick should be imposed in the early phase of the explosion.

Recently ALMA observation found an infrared excess source in the remnant of SN 1987A \citep{cigan19, page20}. 
Assuming that there is a cooling NS at the peak of the infrared excess, 
the NS kick velocity is estimated as $\sim 700\, {\rm km \, s}^{-1}$ from the deviation between the locations of the NS and the progenitor star \citep{cigan19}. 
Although this estimation involves large uncertainties, 
the high kick velocity implies that SN1987A was highly asymmetric in the very beginning of shock evolution. 

\begin{figure*}
  \begin{center}
  \begin{tabular}{cc}
    \includegraphics[width=0.45\linewidth]{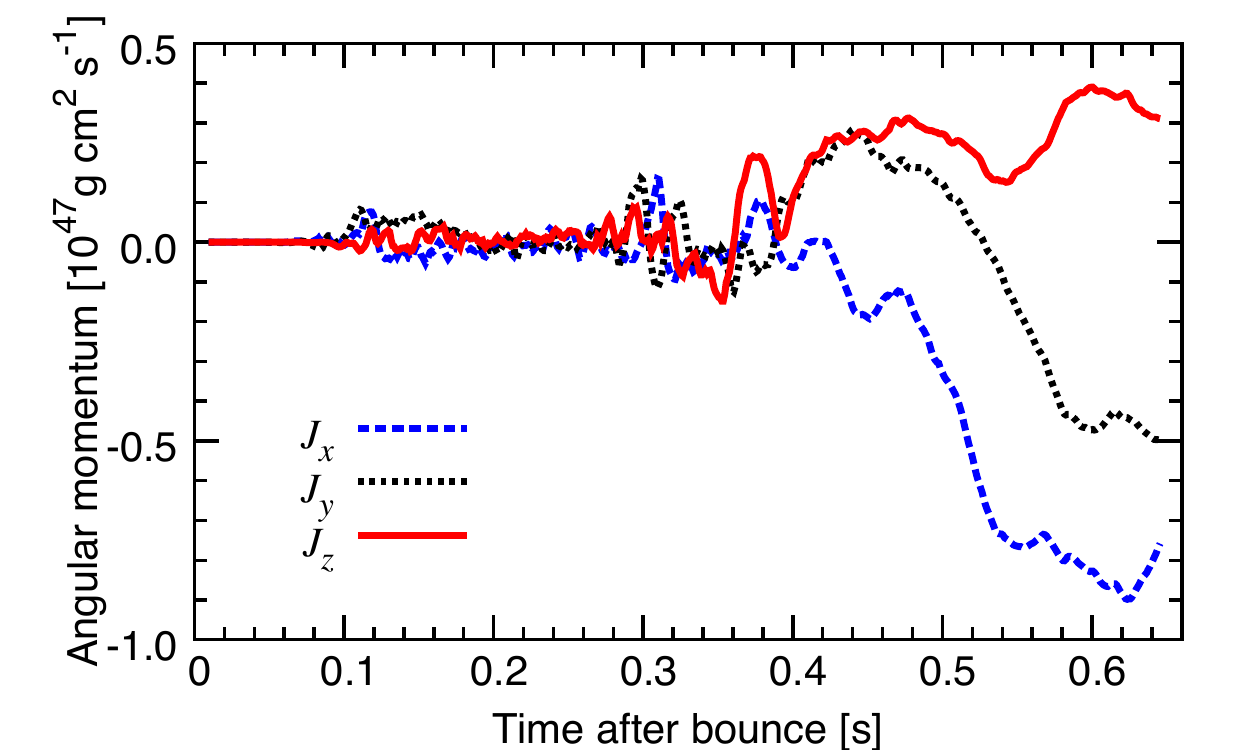} &
    \includegraphics[width=0.45\linewidth]{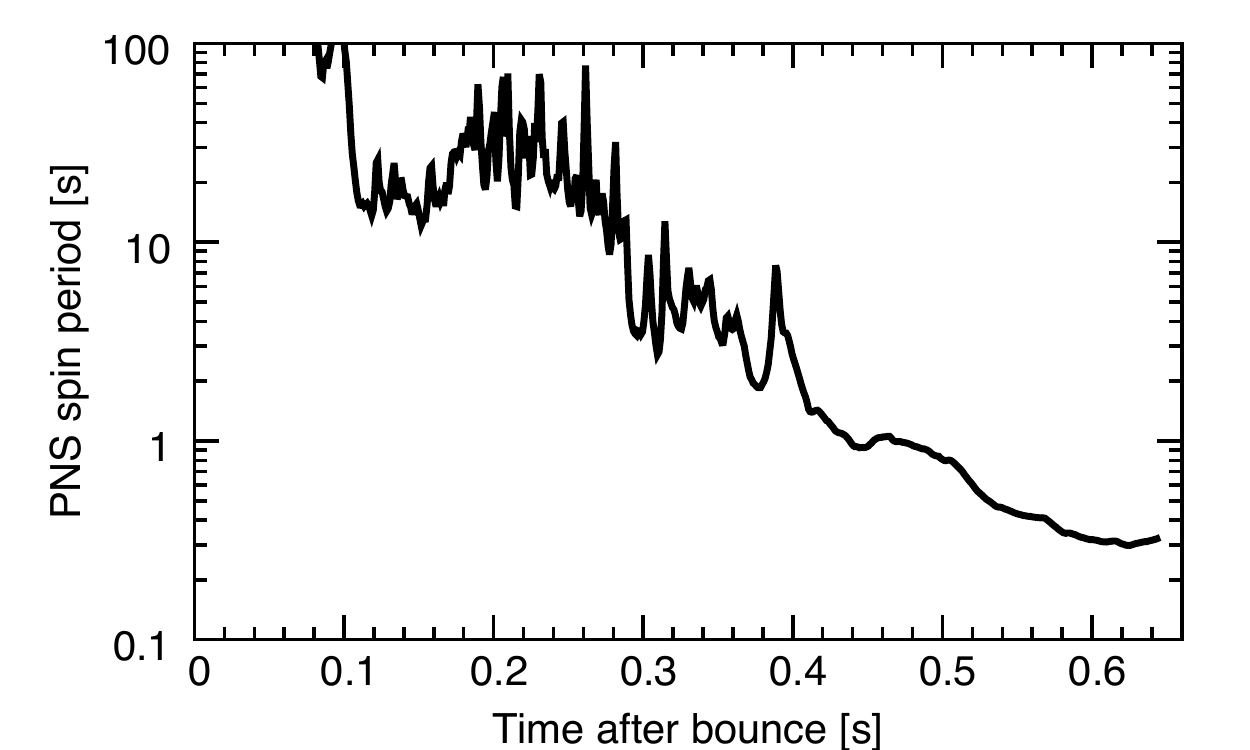}
  \end{tabular}
  \end{center}
  \caption{Left: Angular momentum advected onto the PNS calculated according to Equation \ref{eq:ang_mom}. 
  Right: Evolution of the PNS spin period assuming the moment of inertia of cold neutron stars given by Equation \ref{eq:mom_inrt}.
  }
  \label{fig:pns-spin}
\end{figure*}

Our simulations start from spherically symmetric structure without rotation. 
The non-radial motions developed in our simulations, however, causes the rotation of the PNS in the final state.
We calculate its angular momentum $\mathbf{J}$ in a Newtonian manner 
by integrating the flux of angular momentum through a sphere of radius $r=100$\,km around the origin\footnote{Impact of asymmetric neutrino emission on the kick is disregarded here for simplicity 
\citep[see, however,][]{scheck06,nordhaus10,nagakura19_kick}.}: 
\begin{equation}\label{eq:ang_mom}
    \frac{d\mathbf{J}}{dt} = \int \rho v_r \mathbf{v} \times \mathbf{r} \,d\Omega.
\end{equation}
The estimated angular momentum advected onto the PNS is shown in the left panel of Figure~\ref{fig:pns-spin}. 
The angular momenta around the Cartesian axis fluctuate around zero 
before the stalled shock revive at $\sim 300$\,ms after bounce 
and then develop to $\sim 10^{46-47} {\rm g \, cm^2 \, s^{-1}}$ 
at the final time of our simulation. 
Following \citet{wongwathanarat10}, 
we assume total angular momentum $J_{\rm ns}$ ($= |\mathbf{J}| $) 
and gravitational neutron star mass $M_{\rm grav}$ ($=1.53\,\Msun$ at 0.66\,s after bounce) 
are constant after the end of our simulations. 
We obtain a rough estimate of the NS spin period 
from $T_{\rm ns} = 2 \pi \, I_{\rm ns} /J_{\rm ns}$ 
by considering a moment of inertia $I_{\rm ns}$ given by 
\begin{equation}\label{eq:mom_inrt}
    I_{\rm ns} = 0.237 M_{\rm grav} R^2 
    \left[ 1+4.2 \left( \frac{M_{\rm grav} / \Msun}{R /{\rm km}} \right) 
           + 90  \left( \frac{M_{\rm grav} / \Msun}{R /{\rm km}} \right)^4 \right].
\end{equation}
The time evolution of $T_{\rm ns}$ is shown in the right panel of Figure~\ref{fig:pns-spin}.
The PNS spins up during the simulation time and we find $T_{\rm ns} \sim 0.3$\,s 
at the end of our simulation. 
The hot PNS gradually shrinks by energy loss via radiation. 
We obtain the final neutron star spin period $T_{\rm ns} = 106$\,ms with a final neutron star radius $R=12$\,km. 
This spin period is larger than millisecond pulsars spinning up by accretion from a companion star, but consistent with rotation-powered pulsars whose spin periods are below $\sim 1$\,s \citep[][and references therein]{enoto19}.

\section{Summary and discussions} \label{sec:summary}
We have evolved 
the binary progenitor model of SN 1987A \citep{urushibata18} in 3D with the neutrino-hydrodynamics code 3DnSNe. 
The considered stellar model is constructed as a SN 1987A progenitor based on the slow merger scenario 
and it satisfies most of the observed features like the surface N/C ratio, timing of the blueward evolution of $2 \times 10^4$\,yr before the collapse, and the red-to-blue evolution. 

We have found that a shock formed after the core bounce is once stalled at $r \sim 150$\,km, 
and then effective neutrino heating drives anisotropic matter motions behind the shock. 
The ratio of the advection timescale to the neutrino heating timescale exceeds unity at $\sim 200$\,ms and the shock turns to expand at 220\,ms after bounce. 
The expanding shock is roughly spherical, but the matter behind the shock is characterized by 
a few large hot bubbles going outward.

We have also found that anisotropic motions develop after $\sim 150$\,ms post bounce in and out of the PNS.
The outer anisotropic motion appears at $\sim 80$\,km and gradually spreads outward.
At $\sim 260$\,ms it merges with post-shock turbulence and the stalled shock turns to expand. 
The inner anisotropic motion emerges at $\sim 20$\,km 
and it helps transport of neutrinos diffused from the central core of the PNS. 
Specifically, it enhances the heavy-lepton neutrino luminosity compared to the 1D model without anisotropy. 
On the other hand, the luminosity of electron-type neutrino ($\nue$ and $\nub$) is lower in 3D than in 1D, mainly caused by successful shock revival and suppression of accretion onto the central region in the 3D model. 

In our supernova model, we have observed non-spherical lepton-number flux called LESA. 
The multipole ($l \geq 3$) moments evolve first and shift to the dipole and quadrupole moments. This is in line with the findings by
\citet{glas19b}, who identified such asymmetry of lepton-number emission and found a similar time-evolution 
for 9 and 20 $\Msun$ progenitors using different neutrino transport schemes. 

We have examined the detectability of neutrino and GW signals assuming the distance to the SN event to be 10\,kpc.
Super-Kamiokande can detect enough numbers of anti-electron neutrino 
to determine the time of core bounce as well as the time of shock revival.
Catching GW signals from CCSNe is rather difficult but the current detectors (KAGRA, advanced LIGO, and advanced Virgo) could detect not only the strong high-frequency ($> 500$~Hz) component 
but also the relatively weak low-frequency (200--300~Hz) signals from the PNS convection.
Our SN 1987A model leaves a PNS with baryonic mass of $1.53\,\Msun$, a kick velocity of $\sim 70 \, {\rm km \, s}^{-1}$, and a spin period of $\sim 0.1$ s. 
Although the central remnant of SN 1987A is not yet confirmed, these PNS properties are in the range of typical NSs. 

Our self-consistent 3D simulation demonstrates a neutrino-driven shock revival, 
but the explosion with a diagnostic explosion energy $\sim 0.14 \times 10^{51}$\,erg and $< 0.01\,\Msun$ of ejected $^{56}$Ni is very weak compared to the observed properties ($1.5 \times 10^{51}$\,erg and $0.07\,\Msun$).
Although 3D CCSN models usually suffer from such a kind of weak explosion problem, \citet{bollig21} reported that 
they have attained $\sim 10^{51}$\,erg explosion energy and $0.087\,\Msun$ Ni mass by means of a modern 3D supernova simulation for the first time. 
They initiated simulations from the non-spherical progenitor model 
\citep[see, also][]{yoshida19,yadev20,yoshida21,varma21,fields21} 
which is expected to favor large-scale instabilities, and the cycle of continuous downflows 
and outflows of neutrino-heated matter even after the shock revival led to a monotonic rise of the explosion energy from $\sim 10^{50}$\,erg at 0.6\,s, as small as our 3D model, up to $\sim 10^{51}$\,erg at 7\,s after bounce. 
On the other hand, our simulations start from a spherical progenitor. In our 3D model, 
small-scale convective flows dominate the matter motions behind the shock. It results in weak accretion flows behind the shock and a small growth rate of the explosion energy. 
As shown in Figure \ref{fig:eexp_mni-2d3d}, our 2D model presents a more energetic explosion and 
produces 4 times more Ni than the 3D model. The 2D model shows a sloshing mode of SASI and 
the shock is highly aspherical. It leads to an early and vigorous shock revival to the polar 
direction as well as continuous strong downflows to the proto-NS from the equatorial plane.

There are many works of multi-dimensional CCSN simulations targeted to SN~1987A based on single BSG progenitors \citep[e.g.,][]{kifo00, kifo06,wongwathanarat13,wongwathanarat15,utrobin15,utrobin19} and on binary-merger progenitors \citep{ono20,utrobin21}. 
These simulations were performed until minutes, hours, and days so that they can investigate nucleosynthesis and matter mixing in the ejecta. 
Many important findings consistent with observations, such as the outward mixing of newly formed iron-group elements and inward mixing of hydrogen envelope by Rayleigh-Taylor instabilities at composition interfaces, fast-moving nickel clumps, the distinctive lightcurve of SN~1987A, and implications for the NS kick and spin, 
have been achieved. 
Please be aware that the current simulations presented in this paper are intrinsically different from these previous studies. 
In their simulations, the PNS is excised so that the CFL condition is relaxed, which enables 
them to perform long-term simulations. 
In compensation for the simplification, they can not obtain the neutrino property (luminosity 
and energy distribution). Therefore, they had to assume the PNS contraction rate or the neutrino 
property itself so that at least one of the observational constraints of SN 1987A (for example, 
explosion energy) is satisfied. 
The shortage of the explosion energy and Ni mass in our simulations is suggestive. 
Long-term simulations including several additional ingredients which are not considered in our simulations 
such as rotation, magnetic fields, more elaborate neutrino opacity, or even radioactive particle decays like axions \citep{lucente21,caputo21,caputo,mori22} 
might produce more energetic 3D CCSN models 
and help bridging a gap between the observed values and theoretical predictions, which we leave as our future work.

\section*{Acknowledgements}
We thank T. Urushibata and H. Umeda for kindly providing us with the binary progenitor model of SN~1987A examined in this paper. We also thank H. Nagakura for useful discussion. 
This study was supported in part by Grants-in-Aid for Scientific Research of the Japan Society for the Promotion of Science (JSPS, Nos. 
JP20K03939, 
JP21H01121, 
JP21H01088, 
JP22H01223), 
the Ministry of Education, Science and Culture of Japan (MEXT, Nos. 
JP17H06364, 
JP17H06365, 
JP19H05811, 
JP20H04748, 
JP20H05255), 
 by the Central Research Institute of Explosive Stellar Phenomena (REISEP) at Fukuoka University and an associated project (No. 207002),
and JICFuS as “Program for Promoting researches on the Supercomputer Fugaku” (Toward a unified view of 
the universe: from large scale structures to planets, JPMXP1020200109).
Numerical computations were in part carried out on Cray XC50 at Center for Computational Astrophysics, National Astronomical Observatory of Japan.

\section*{Data Availability}

The data underlying this article will be shared on reasonable request to the corresponding author.

\bibliographystyle{mnras}
\bibliography{reference}

%
%
%



\appendix

\bsp	
\label{lastpage}
\end{document}